\documentclass[acmlarge,screen]{acmart}
\usepackage[ruled]{algorithm2e}
\usepackage{graphicx}
\usepackage{textcomp}
\usepackage{xparse,mleftright}
\usepackage{threeparttable}
\usepackage{diagbox}
\usepackage{graphicx}
\usepackage{textcomp}
\usepackage{xcolor}
\usepackage{url}
\usepackage{listings}
\usepackage{graphicx}
\usepackage{float} 
\usepackage{multirow}
\usepackage{subfigure}
\usepackage{bm}
\usepackage{fancyhdr}
\usepackage{tcolorbox}
\usepackage{booktabs}
\usepackage{colortbl}
\usepackage{bbding}
\usepackage{pifont}
\usepackage{color}
\usepackage[noend]{algpseudocode}

\usepackage{enumitem}
\usepackage{microtype}
\usepackage{makecell}
\usepackage{amsmath} 
\usepackage{minted}

\AtBeginDocument{%
  }

\setcopyright{acmcopyright}
\copyrightyear{2025}
\acmYear{2025}
\acmJournal{TOSEM}
\acmDOI{XXXXXXX.XXXXXXX}
\received{2024-10-12}
\received[accepted]{2025-04-02}

\begin{document}

\title{A Survey on Failure Analysis and Fault Injection in AI Systems}

\author{Guangba Yu}
\email{yugb5@mail2.sysu.edu.cn}
\orcid{0000-0001-6195-9088}
\author{Gou Tan}
\email{tang29@mail2.sysu.edu.cn}
\orcid{0009-0008-6580-1470}
\author{Haojia Huang}
\orcid{0009-0008-5592-7782}
\email{huanghj78@mail2.sysu.edu.cn}
\affiliation{%
  \institution{Sun Yat-sen University}
  \city{Guangzhou}
  \country{China}
}

\author{Zhenyu Zhang}
\orcid{0009-0002-5130-0567}
\email{zhangzhy239@mail2.sysu.edu.cn}
\author{Pengfei Chen}
\orcid{0000-0003-0972-6900}
\email{chenpf7@mail.sysu.edu.cn}
\affiliation{%
  \institution{Sun Yat-sen University}
  \city{Guangzhou}
  \country{China}
}

\author{Roberto Natella}
\orcid{0000-0003-1084-4824}
\email{roberto.natella@unina.it}
\affiliation{%
  \institution{Federico II University of Naples}
  \city{Naples}
  \country{Italy}
}

\author{Zibin Zheng}
\orcid{0000-0002-7878-4330}
\email{zhzibin@mail.sysu.edu.cn}
\affiliation{%
  \institution{Sun Yat-sen University}
  \city{Zhuhai}
  \country{China}
}

\author{Michael R. Lyu}
\orcid{0000-0002-3666-5798}
\email{lyu@cse.cuhk.edu.hk}
\affiliation{%
  \institution{Chinese University of Hong Kong}
  \city{Hong Kong}
  \country{China}
}


\newcommand\papernum{142\xspace}


\renewcommand{\shortauthors}{Guangba Yu, Gou Tan, Haojia Huang, Zhenyu Zhang \textit{et al.}}

\begin{abstract}
The rapid advancement of Artificial Intelligence (AI) has led to its integration into various areas, especially with Large Language Models (LLMs) significantly enhancing capabilities in Artificial Intelligence Generated Content (AIGC). However, the complexity of AI systems has also exposed their vulnerabilities, necessitating robust methods for failure analysis (FA) and fault injection (FI) to ensure resilience and reliability. Despite the importance of these techniques, there lacks a comprehensive review of FA and FI methodologies in AI systems. This study fills this gap by presenting a detailed survey of existing FA and FI approaches across six layers of AI systems. We systematically analyze \papernum studies to answer three research questions including (1) what are the prevalent failures in AI systems, (2) what types of faults can current FI tools simulate, (3) what gaps exist between the simulated faults and real-world failures. Our findings reveal a taxonomy of AI system failures, assess the capabilities of existing FI tools, and highlight discrepancies between real-world and simulated failures. Moreover, this survey contributes to the field by providing a framework for fault diagnosis, evaluating the state-of-the-art in FI, and identifying areas for improvement in FI techniques to enhance the resilience of AI systems.
\end{abstract}




\maketitle

\section{Introduction}

Artificial intelligence (AI) has made significant strides in the past decade, permeating both academic and industrial areas. Large Language Models (LLMs), in particular, have proven to be a game changer, propelling AI to unprecedented heights and facilitating a myriad of applications in fields such as software engineering~\cite{LLMSE1,LLMSE2,LLMSE3} and human language translation~\cite{kocmi2023large,hendy2023good,bang-2023-gptcache}. This evolution has led to the integration of AI models into a growing array of products, which transform them into sophisticated AI systems. Notable examples of such integration include Gemini~\cite{gemini}, Bing~\cite{bingai}, and ChatGPT~\cite{chatgpt}, which underscore the pivotal role of AI in enhancing and expanding the capabilities of modern technology solutions.

The increasing complexity and ubiquity of AI systems necessitate addressing their inherent vulnerabilities and failure-related challenges. A Meta AI report~\cite{OPTSusan2022} points out more than 100 failures during OPT-175B training. Similarly, ChatGPT encountered 173 outages in 2023, causing a maximum user impact of 427 minutes~\cite{chatgptincident}. Such failures can degrade user experience and even incur financial losses. Hence, mitigating AI system failures is of paramount importance.

The real-world implications of failure analysis in AI systems are far-reaching. They not only affect the performance and reliability of AI systems, but also have a profound impact on the industries and sectors that rely on these systems. For example, in healthcare, a failure in an AI system could lead to incorrect diagnoses or treatment recommendations, potentially endangering lives. In the financial sector, AI system failures could result in significant financial losses due to erroneous transactions or predictions. Therefore, understanding and addressing these failures is not just a technical challenge but a social imperative.

Failure analysis (FA) and fault injection (FI) techniques are instrumental in identifying limitations and bolstering the reliability of AI systems. Researchers and practitioners alike have conducted extensive investigations into AI system failures. Studies~\cite{LiuRise2023,ChenToward2023,ZhangAn2018,TarekAn2022,FlorianSilent2024,ZengyangUnderstanding2023,IslamComprehensive2019,HumbatovaTaxonomy2020} have analyzed AI system failures from platforms such as Stack Overflow or Github, while others~\cite{JeonAnalysis2019,ZhangEmpirical2020,EmpiricalGao2023,OstrouchovGPU2020,FailuresGupta2017,ExaminingTaherin2021,TiwariReliability2015} have focused on failures in large-scale production AI systems. Such failure analyses enable the identification of patterns, root causes, and locations, thereby informing FI techniques. FI, a proactive approach, uncovers system weaknesses in resilience before they become catastrophic failures. By deliberately injecting faults or abnormal conditions into systems, teams can evaluate and enhance their resilience to unexpected disruptions. Some existing FI approaches~\cite{AbrahamFault2022,ChanUnderstanding2021,LeiDeepMutation2018,SambasivanEveryone2021,DanBenchmarking2019,LiHow2022,HumbatovaDeepCrime2021,ShenMuNN2018} mimic faults in AI systems engineered by humans, while others~\cite{ZhaoQuantifying2019,GuanpengUnderstanding2017,DanielRadiation2017,BehzadResilience2018,JeffFault2019,JiachaoRetraining2015} simulate hardware errors.

\begin{figure}[t]
  \centering
  \begin{minipage}{0.4\textwidth}
    \centering
    \includegraphics[width=\linewidth]{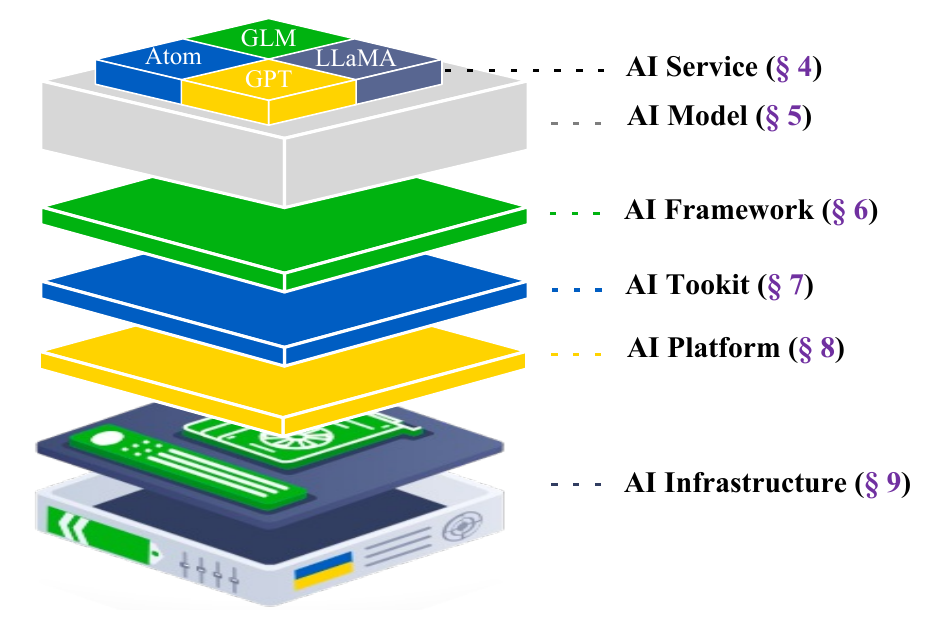}
    \vspace{-0.1in}
    \caption{An overall framework of AI system.}
    \label{fig:layer}
  \end{minipage}
  \hfill
  \begin{minipage}{0.59\textwidth}
    \centering
    \captionof{table}{Definition of layers in AI system.}
    \vspace{-0.1in}
    \resizebox{\linewidth}{!}{
\begin{tabular}{@{}lp{7cm}c@{}}
\toprule
\textbf{Stack}             & \textbf{Definition}                                                                                                & \textbf{Representative} \\ \midrule
\multirow{2}{*}{AI Service}         & Well-defined AI-enabled services that can be fulfilled by API function call request.                         & \multirow{2}{*}{ChatGPT}        \\\hline
\multirow{2}{*}{AI Model}          & AI-related models that power the AI capabilities of the system                                               & \multirow{2}{*}{GPT-4}          \\\hline
\multirow{3}{*}{AI Framework} & AI-related frameworks for defining AI/ML architectures and invoking algorithms, library, and accelerator drive on the hardware. & \multirow{3}{*}{Pytorch} \\\hline
\multirow{2}{*}{AI Tookit}         & An intermediate interface between the AI framework and underlying devices                                    & \multirow{2}{*}{CUDA}           \\\hline
\multirow{3}{*}{AI Platform}       & AI-related platform that enable large-scale organizations to design, develop, deploy and operate AI services & \multirow{3}{*}{Ray}            \\\hline
\multirow{2}{*}{AI Infrastructure} & Hysical and virtualized resources necessary for the deployment and operation of AI services.                 & \multirow{2}{*}{GPU, TPU}       \\ \bottomrule
\end{tabular}%
}
    \label{tab:layer}
  \end{minipage}
\vspace{-0.25in}  
\end{figure}

Despite progress, a comprehensive survey on FA and FI in AI systems is conspicuously absent. Furthermore, there is a gap between FI and FA, resulting in insufficient consideration of FA outcomes when designing FI tools. Therefore, this study presents a comprehensive survey aimed at exploring and evaluating existing research on FA and FI in AI systems. We meticulously reviewed and analyzed \papernum corresponding publications. Motivated by previous studies~\cite{TsaihAI2023}, we divide an AI system into six layers including AI Service, AI Model, AI Framework, AI Toolkit, AI Platform, and AI Infrastructure. The relationships between these layers are illustrated in Fig.~\ref{fig:layer}, while detailed explanations are provided in Table~\ref{tab:layer}. We attempt to address three research questions at each layer as follows.



\begin{itemize}
\item \textbf{RQ1:} What are the prevalent failures in current AI systems?
\item \textbf{RQ2:} What types of fault can current FI tools simulate?
\item \textbf{RQ3:} What gaps exist between simulated faults and real-world failures?
\end{itemize}

\textbf{RQ1} aims to catalog and analyze the failures that have occurred in current AI systems. Understanding these failures is crucial for several reasons. Since it helps to identify common vulnerabilities within AI systems, it informs developers about potential areas of improvement and contributes to the development of more reliable AI applications.
\textbf{RQ2} explores the capabilities of existing FI tools designed for AI systems. The ability to simulate a wide range of faults is essential for evaluating and improving the robustness and fault tolerance of AI systems.
\textbf{RQ3} investigates the gap between simulated faults and real-world AI system failures, with the aim of understanding the limitations of current FI tools in producing the full spectrum of potential failures.  Moreover, understanding these gaps helps to improve FI tools and ultimately contributes to develop more resilient AI systems.

By examining the current landscape and identifying critical research gaps, this survey provides valuable insights for researchers and practitioners working toward building reliable and resilient AI systems. This study makes the following contributions.
\begin{itemize}
\item We present a comprehensive analysis and taxonomy of failures that occur in different layers of AI systems. By systematically characterizing these failures, we provide a valuable framework that can serve as a reference for failure diagnosis in AI systems. This taxonomy paves the way for future research by highlighting specific areas where AI systems are most vulnerable, thereby guiding targeted improvements and innovations.
\item We conduct an in-depth examination of the capabilities of existing FI tools across various layers of AI systems. We offer insights into the state-of-the-art in simulating and reproducing potential failures. This work provides a foundation for assessing the reliability of AI systems. Our findings underscore the importance of developing more sophisticated FI tools that can better mimic real-world failure scenarios, thus enhancing the robustness of AI systems.
\item We explore the discrepancies between FI tools and real-world AI system failures. We identify the limitations of current FI approaches in simulating potential failure scenarios. By shedding light on these gaps, we emphasize the need for more comprehensive FI techniques in AI systems. Addressing these gaps is crucial for advancing the field, as it will lead to the creation of more resilient AI systems capable of withstanding a broader range of failure conditions.
\end{itemize}

The remainder of this paper is organized as follows. Section~\ref{sec:bak} provides background information on FA and FI in AI systems, followed by Section~\ref{sec:method}, which outlines our systematic literature review methodology. The subsequent sections analyze FA and FI in different layers of AI systems, including the AI service layer (Section~\ref{sec:service}), AI model layer (Section~\ref{sec:model}), AI framework layer (Section~\ref{sec:framework}), AI toolkit layer (Section~\ref{sec:toolkit}), AI platform layer (Section~\ref{sec:platform}), and AI infrastructure layer (Section~\ref{sec:infrastructure}). 
Section~\ref{sec:threats_to_validity} outlines potential threats to validity in our study.  
Section~\ref{sec:opportunities} highlights research opportunities on FI in AI systems. 
The article concludes in Section~\ref{sec:conclusion}.





\section{Background and definitions}\label{sec:bak}

\subsection{Failures and Faults}
We adopt the definitions of failures and faults proposed in previous work~\cite{Fault04TDSC,SurveyFelix2010}. Furthermore, we provide additional extensions and interpretations specific to AI systems.
\begin{itemize}[leftmargin=*]
\item \textbf{Failure} is defined as ``an incident that occurs when the delivered service deviates from the correct service'' ~\cite{Fault04TDSC}. In the context of AI systems, failures can manifest in various ways. For example, a failure can occur when AI services become unreachable and when the behavior of AI services does not meet the expected outcome (e.g., generating semantically incorrect text). These failures indicate a deviation from the desired or expected behavior of the AI system.
\item \textbf{Fault} is the  root cause of a failure. In AI systems, faults can be attributed to various sources, including algorithmic flaws, model design issues, or problems with the quality of the data used for training or inference. It is important to note that faults in AI systems may remain uncovered for some time, due to fault-tolerant approaches implemented in the system. 
\end{itemize}

\subsection{Failure Comparisons between AI and Cloud System}

Failure analysis (FA) and fault injection (FI) are longstanding topics within the field of computer science, traditionally focusing on the robustness and reliability of systems. Historically, much of the literature has focused on cloud systems, reflecting their critical role in modern computing infrastructure~\cite{Supriyoincidents2022SoCC,XiaoyunFaults2022ISSRE,Michaelpostmortems2016ICSE,HaryadiOutages2016SoCC,Haopengbugs2019HotOS,JonathanFailures2020ICSME}. These systems adhere to a logic-based programming paradigm, where developers encode decision logic directly into the source code, facilitating a structured approach to FA and FI. In contrast, AI systems represent a paradigm shift towards a data-driven programming model. Here, developers design neural networks that derive decision-making logic from extensive datasets~\cite{LiuRise2023,ChenToward2023,ZhangAn2018,TarekAn2022,FlorianSilent2024,ZengyangUnderstanding2023,IslamComprehensive2019,HumbatovaTaxonomy2020}. This shift introduces both similarities and differences in the approach to FA between AI systems and traditional cloud systems.

\begin{figure}[t]
\centering
\includegraphics[width=\linewidth]{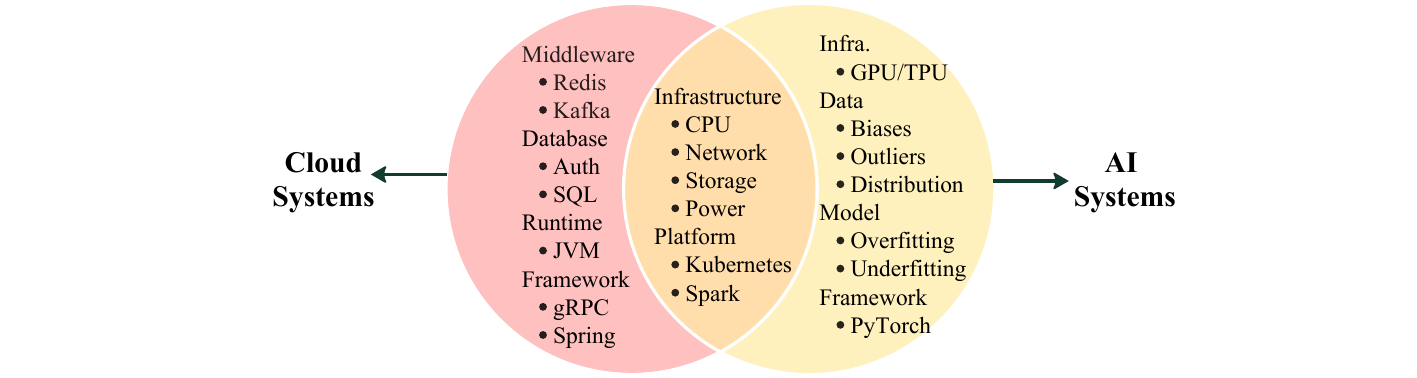} 
\vspace{-0.20in}
\caption{Failure comparisons between AI and cloud systems.}
\vspace{-0.1in}
\label{fig:difference}
\end{figure}

As illustrated in Fig.~\ref{fig:difference}, while AI and cloud systems are susceptible to common failures such as power disruptions and network outages, certain faults are unique or more critical to AI systems. For instance, GPU failures, which might be relatively inconsequential in traditional cloud environments, can severely affect the performance and availability of AI systems. This distinction underscores the importance of conducting a comprehensive survey on FA and FI specifically for AI systems, especially in the current era dominated by ``Large Models''. This need forms one of the primary motivations behind the research presented in this article. 

\subsection{Failure Analysis and Fault Injection}

\begin{figure}[t]
\centering
\includegraphics[width=0.8\linewidth]{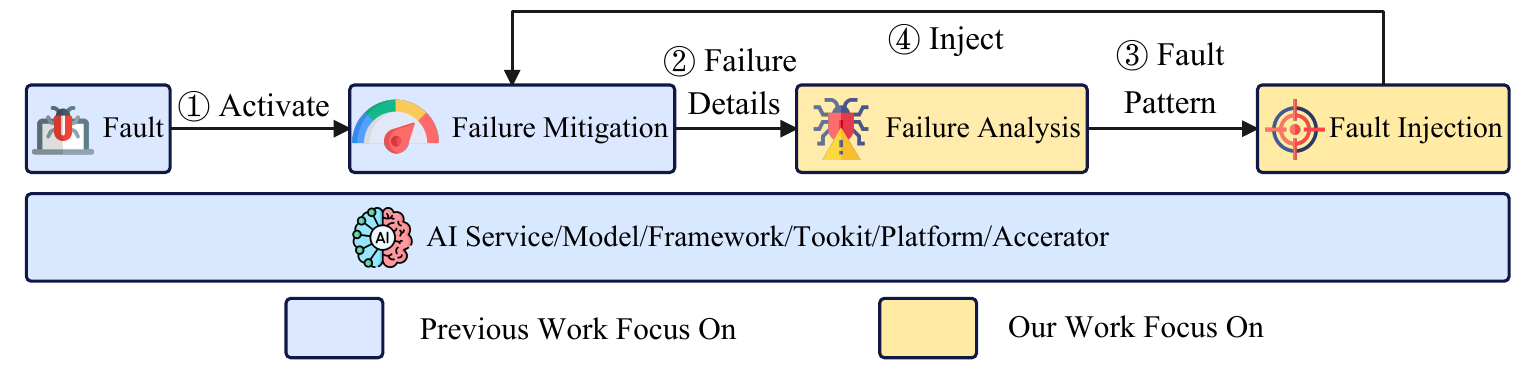} 
\vspace{-0.2in}
\caption{Overall life cycle of an AI system failure.}
\vspace{-0.1in}
\label{fig:process}
\end{figure}

Figure~\ref{fig:process} demonstrates the relationship between faults, FA, and FI. In large-scale AI systems, faults commonly occur in AI systems due to their inherent complexity and the numerous interconnected components involved, leading to fault origination at various stages of system operation~\cite{LiuRise2023,ChenToward2023,ZhangAn2018,TarekAn2022,FlorianSilent2024,IslamComprehensive2019,HumbatovaTaxonomy2020}.
Once a fault is activated and a failure is detected (stage \textcircled{\raisebox{-0.9pt}{1}}), engineers responsible for AI system maintenance engage in mitigating the failure based on the observed behaviors (stage \textcircled{\raisebox{-0.9pt}{2}}). During the failure mitigation process, engineers generate comprehensive incident reports encompassing failure details such as occurrence time, impact, failure manifestation, and mitigation strategies. The objective of FA is to utilize these incident reports as inputs to summarize the fault pattern (e.g., recurring type and location) (stage \textcircled{\raisebox{-0.9pt}{3}}).

FI is a widely adopted technique to assess and improve the reliability and security of systems, including AI systems. It involves the deliberate introduction of faults into a system to observe its behavior and validate its fault tolerance mechanisms. FI can be applied in various forms, such as software fault injection (e.g., mutation test~\cite{JiaAn2011} and API interception~\cite{nezha,WanAPIs21}) or hardware fault injection (e.g., simulating hardware failures~\cite{GPUDSN2018} and environmental disturbances~\cite{Hassan23Network}).

Leveraging the knowledge acquired from the historical FA, engineers employ FI techniques to validate the reliability of AI systems.
Following injection of faults, engineers closely monitor AI system performance and behavior, ensuring the accurate identification and appropriate handling of injected faults (stage \textcircled{\raisebox{-0.9pt}{4}}). 
Based on the analysis of FI experiments, engineers can identify system vulnerabilities, weaknesses, and areas for improvement. By iteratively conducting FI experiments and refining the system based on the results obtained, the AI system can be continually improved, enhancing its reliability and effectiveness in real-world scenarios.


Consequently, FA and FI form closely intertwined processes that contribute significantly to the evaluation and enhancement of AI system reliability. The insights derived from FA guide the selection and design of FI scenarios. The iterative feedback loop established by fault analysis and FI facilitates the continuous improvement of AI systems, thereby serving as the other motivation driving the research presented in this article.

\section{Survey Methodology}\label{sec:method}

\begin{figure}[t]
\centering
\includegraphics[width=0.9\linewidth]{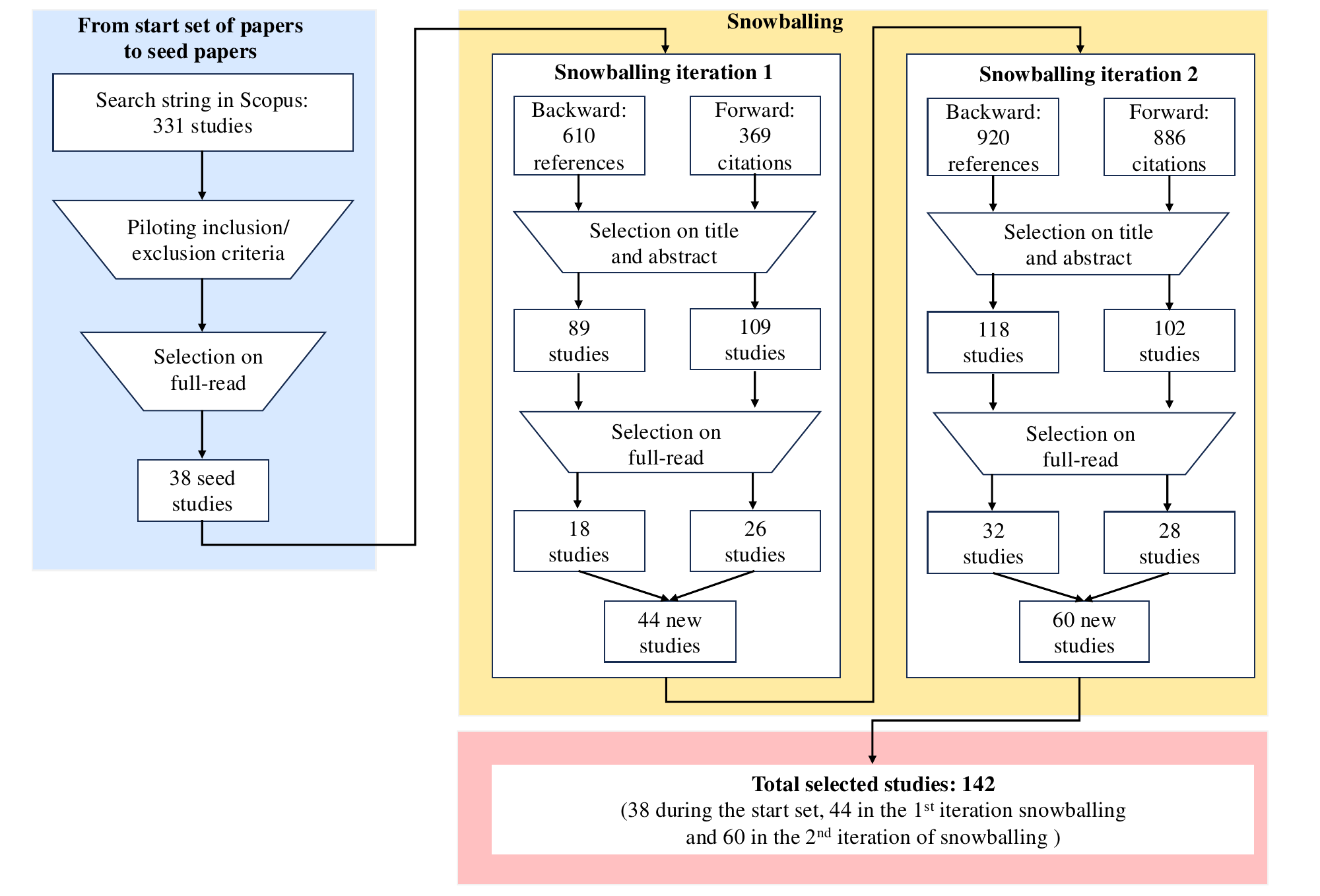} 
\vspace{-0.1in}
\caption{Search strategy of our study.}
\vspace{-0.1in}
\label{fig:method}
\end{figure}

\subsection{Search Strategy}\label{sec:search_strategy}

We visualize our research process in Fig.~\ref{fig:method} and describe the different steps in this section. Our study employs a hybrid search strategy~\cite{search}, combining a systematic search in Scopus~\footnote{https://www.scopus.com/} with a bidirectional snowballing approach (referred to as ``Scopus + BS*FS'')~\cite{snowballing}. This strategy first executes a search query in Scopus to obtain an initial set of papers, which serves as the seed set for subsequent snowballing. Scopus was chosen as the primary database due to its extensive coverage of peer-reviewed publications from top software engineering journals and conferences, including IEEE Xplore, ACM Digital Library, ScienceDirect (Elsevier), and Springer. Previous studies~\cite{scopus} have demonstrated that Scopus provides optimal coverage compared to other databases. 

The search string was iteratively refined through discussions among the authors, ensuring it captured relevant studies while minimizing noise. The final search string executed in September 2024 was as follows:

\begin{center}
\begin{minipage}{0.9\textwidth}
\small 
\begin{verbatim}
( TITLE (Failure OR Fault OR "Fault Injection" OR "Bug" OR "Error" ) 
AND 
TITLE-ABS-KEY ( "AI System" OR "AI Application" 
OR "Artificial Intelligence System" OR "Artificial Intelligence Application" 
OR "ML System" OR "ML Software" 
OR "Machine Learning System" OR "Machine Learning Application" 
OR "Deep Learning System" OR "Deep Learning Application" 
OR "DL System" OR "DL Application"
OR "Large Language Model System" OR "Large Language Model Application" 
OR "LLM System" OR "LLM Application")
AND
PUBYEAR > 1999 AND PUBYEAR < 2025 
AND 
(LIMIT-TO ( SUBJAREA , "COMP" ) )
\end{verbatim}
\end{minipage}
\end{center}

This query yielded 331 potentially relevant studies. Each study was independently screened by two researchers using predefined inclusion and exclusion criteria (detailed in Section~\ref{sec:screening}). After screening, 38 studies formed the seed set for snowballing. We conducted two rounds of bidirectional snowballing (Section~\ref{sec:snowballing}), resulting in a final set of \papernum publications on failure analysis (FA) and fault injection (FI) in AI systems, spanning the years 2001 to 2024.

\subsubsection{Screening of Studies}\label{sec:screening}

The following inclusion (I) and exclusion (E) criteria were applied during the selection of primary studies:
\begin{itemize}
    \item \textbf{Inclusion criteria}
    \begin{itemize}
        \item \textbf{IC1}: The study primarily focuses on failure analysis for AI systems.
        \item \textbf{IC2}: The study primarily focuses on fault injection for AI systems.
    \end{itemize}
    \item \textbf{Exclusion criteria}
    \begin{itemize}
        \item \textbf{EC1}: The study does not fulfill IC1 or IC2.
        \item \textbf{EC2}: The study is not accessible, even after contacting the authors.
        \item \textbf{EC3}: The study is not entirely written in English.
        \item \textbf{EC4}: The study is a pure secondary or tertiary study.
        \item \textbf{EC5}: The study was published before January 2000 or after September 2024.
    \end{itemize}
\end{itemize}
A study had to meet all inclusion criteria and none of the exclusion criteria to be included in the primary study set. To ensure consistency, a pilot selection process was conducted, where three researchers with expertise in AI systems and software engineering independently evaluated five studies and discussed their decisions in a consensus meeting. After refining the selection process, it was applied to all potentially relevant studies. Each study was independently evaluated by two researchers, and consensus was required for final inclusion. This process resulted in 38 studies forming the seed set for snowballing.

\subsubsection{Snowballing Approach}\label{sec:snowballing}

To comprehensively identify relevant studies, we applied both backward and forward snowballing, as recommended by Wohlin et al.~\cite{snowballing}. In this process, all papers that either cited or were cited by a paper in the seed set were evaluated for inclusion in the final set of primary studies. The same inclusion and exclusion criteria and screening process were applied, with each study independently evaluated by two researchers and consensus required for inclusion. After the first round of bidirectional snowballing, 44 additional studies were included. A second round of bidirectional snowballing added 60 more studies, resulting in a final set of \papernum publications.

\subsection{Data Extraction and Mapping Process}

To address the research questions (RQ1–RQ3) and systematically analyze the relationships between failure analysis (FA) and fault injection (FI) in AI systems, we developed a structured data extraction and mapping process. This process was designed to ensure consistency, reproducibility, and alignment with the six-layer AI system architecture (AI Service, AI Model, AI Framework, AI Toolkit, AI Platform, and AI Infrastructure) illustrated in Fig.~\ref{fig:layer} and detailed in Table~\ref{tab:layer}.

\subsubsection{Data Extraction Framework}
We defined a data extraction framework to systematically collect and organize relevant information from the \papernum primary studies. The framework included the following key data points for each study:
\begin{itemize}
\item \textbf{Study Metadata}: Publication year, venue, and type (e.g., journal, conference).
\item \textbf{AI System Layer}: The layer(s) of the AI system addressed in the study (e.g., AI Model, AI Service).
\item \textbf{Failure Analysis (FA)}: Types of failures reported, failure causes, and their impact on the AI system.
\item \textbf{Fault Injection (FI)}: Types of faults simulated, FI techniques used, and their alignment with real-world failures.
\item \textbf{Gaps Between FA and FI}: Identified discrepancies between simulated faults and real-world failures, as well as limitations of current FI tools.
\end{itemize}

\subsubsection{Mapping Process}
The extracted data was mapped to the six-layer AI system architecture to address the research questions at each layer. The mapping process involved the following steps:
\begin{enumerate}
\item \textbf{Classification by Layer}: Each study was classified into one or more layers based on its primary focus. For example, a study discussing failures in AI model training was classified under the AI Model layer.
\item \textbf{Iterative Validation}: Two authors independently classified the studies, and disagreements were resolved through consensus meetings. This iterative process ensured accuracy and consistency in the classification.
\item \textbf{Cross-Layer Analysis}: Studies addressing multiple layers were analyzed to identify inter-layer dependencies and interactions. For instance, failures in the AI Service layer were often linked to underlying issues in the AI Model or AI Infrastructure layers.
\end{enumerate}

\subsubsection{Addressing Research Questions}
The extracted and mapped data was analyzed to answer the three research questions:
\begin{itemize}
\item \textbf{RQ1 (Prevalent Failures)}: We cataloged and analyzed failures reported in each layer, identifying common vulnerabilities and their root causes. This analysis provides insights into the types of failures that occur in real-world AI systems and highlights areas for improvement.
\item \textbf{RQ2 (FI Tool Capabilities)}: We evaluated the types of faults simulated by existing FI tools, assessing their coverage and effectiveness in testing AI system robustness. This analysis reveals the strengths and limitations of current FI tools.
\item \textbf{RQ3 (Gaps Between FA and FI)}: We identified discrepancies between simulated faults and real-world failures, uncovering limitations in current FI tools. This analysis informs the design of more effective FI tools that better reflect real-world failure scenarios.
\end{itemize}

\subsubsection{Quality Assurance}
To ensure the reliability and validity of the data extraction and mapping process, we implemented the following quality assurance measures:
\begin{itemize}
\item \textbf{Inter-Rater Reliability}: Two authors independently extracted and mapped data for a subset of studies, and the inter-rater agreement was measured using Cohen’s Kappa coefficient~\cite{cohen1960}.  The inter-rater reliability for classification was measured using Cohen's Kappa (k = 0.82), indicating high consistency. Disagreements were resolved through discussion and consensus.
\item \textbf{Iterative Refinement}: The data extraction framework and mapping process were refined iteratively based on feedback from the authors and preliminary findings. This refinement ensured the framework captured all relevant data points and addressed the research questions effectively.
\end{itemize}

Through this rigorous data extraction and mapping process, we systematically analyzed the relationships between FA and FI in AI systems, providing a comprehensive understanding of the gaps and opportunities for future research.

\subsection{Summary of Our Survey}

\begin{figure}[t]
\centering
\includegraphics[width=0.8\linewidth]{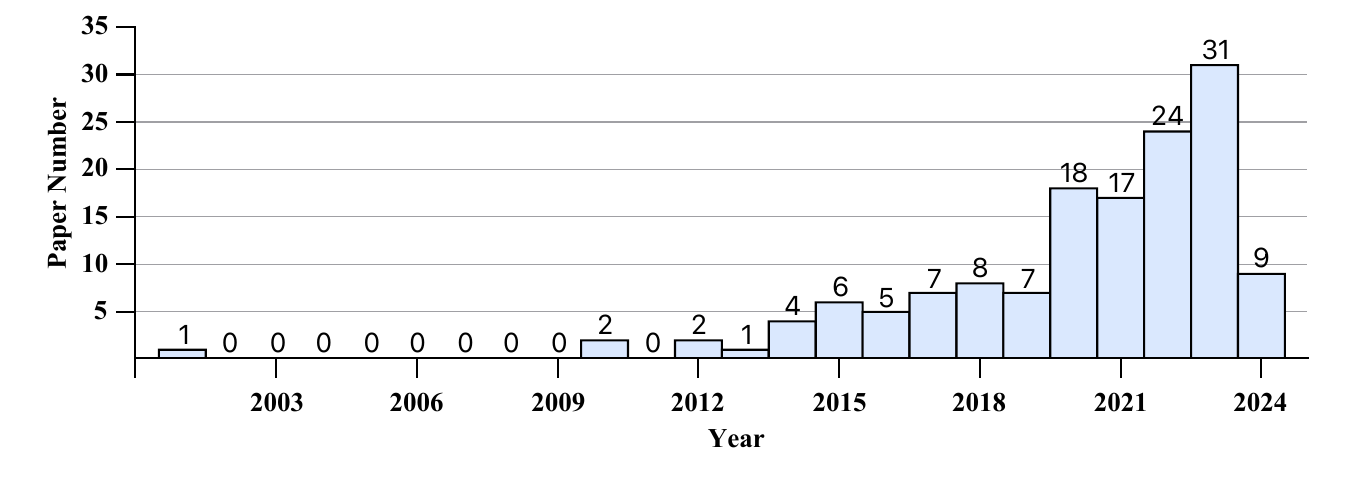} 
\vspace{-0.2in}
\caption{Distribution of studies across layers.}
\vspace{-0.10in}
\label{fig:year}
\end{figure}

In total, our recursive search methodology allowed us to collect \papernum studies on FA and FI in AI systems, covering 2001 to 2024. \autoref{fig:year} shows the histogram of annual studies during the period.  We can FA and FI in AI systems has been continuously and actively investigated in the past decades. In particular, we can observe a steady growth in the number of studies since 2015, indicating that the AI systems field has attracted an increasing amount of interest since then.

\begin{figure}[t]
\centering
\includegraphics[width=0.8\linewidth]{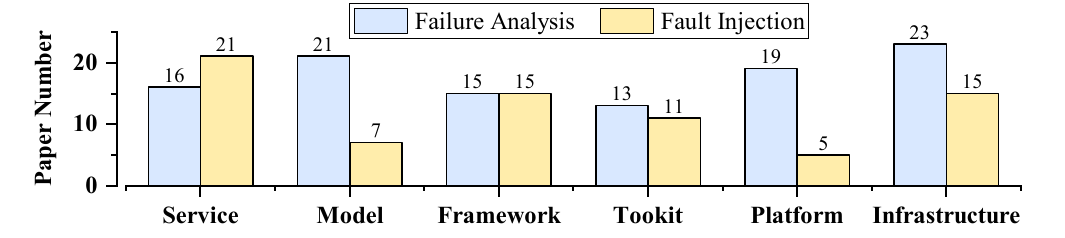} 
\vspace{-0.2in}
\caption{Distribution of failure analysis and fault injection studies across layers.}
\vspace{-0.10in}
\label{fig:paper}
\end{figure}

These studies are classified into six layers by research topics including AI service, AI model, AI framework, AI toolkit, AI platform, and AI infrastructure. Moreover, the distribution of different classes is presented in Fig.~\ref{fig:paper}. It is important to note that some studies may address more than one research topic. Consequently, the total number of studies in Fig.~\ref{fig:paper} is larger than \papernum. Next, we will show the details on FA and FI in different layers of AI systems.

\subsection{Comparing Against Existing Surveys}

There are three existing works~\cite{Natella2016CSUR,Bahar2022Systematic,Mart2022Survey} that are related to our survey.
\begin{itemize}
    \item Silverio~\textit{et al.}~\cite{Mart2022Survey} conducted a survey on the literature in the field of Software Engineering for Artificial Intelligence. They found that the most studied characteristics of AI systems are reliability and security. Research related to software testing and software quality is prevalent, but areas such as software maintenance seem to be overlooked, with fewer related papers. The focus of this paper is on failure analysis, which falls under the umbrella of software maintenance, thereby addressing a gap that was not considered in their article. This highlights the importance and relevance of our survey, as it contributes to a more comprehensive understanding of the challenges and solutions in the field of AI systems. 
    \item Bahar~\textit{et al.}~\cite{Bahar2022Systematic} aimed to investigate the recent advancements in Software Quality (SQ) based on AI systems and identify quality attributes, application models, challenges, and practices reported in the literature. Their focus was primarily on how to develop a high-quality AI system. However, they did not delve into the analysis and handling of AI system failures. Our study complements the work of Bahar~\textit{et al.}~\cite{Bahar2022Systematic} by providing a more comprehensive view of software quality in AI systems, extending beyond development to include maintenance and fault tolerance.
    \item Natella~\textit{et al.}~\cite{Natella2016CSUR} provided a comprehensive overview of the latest advancements in software fault injection to assist researchers and practitioners in selecting the methods most suitable for their reliability assessment goals. However, their focus was primarily on traditional software systems, not AI systems. As shown in Fig.~\ref{fig:difference}, AI systems differ significantly from previous systems. They have unique characteristics and complexities, such as the use of machine learning models, which introduce new types of fault~\cite{deployai,liu2022ISSRE}. Therefore, there is a need for research specifically targeted at AI systems.
\end{itemize}
Therefore, we claim that our study is the first to provide a comprehensive review of the scientific literature in the context of FA and FI specifically for AI-based systems. By focusing on the unique challenges and characteristics of AI systems, our objective is to fill a critical gap in the existing literature and provide valuable insights to researchers and practitioners in this rapidly evolving field. Our survey not only synthesizes the current state of the art, but also identifies key areas for future research, thereby paving the way for the development of more reliable and robust AI systems.






\section{Failure Analysis and Fault Injection in AI Service}\label{sec:service}
In the context of AI systems, the AI service layer can be considered as the top layer that interacts directly with users or other systems. This layer acts as an interface or entry point for users to access and utilize the AI capabilities provided by the underlying layers of the AI system. Potential failures at this layer can include unavailability of the service or incorrect output, such as incorrect inference from a classifier or hallucinations from an LLM-based service. This section delves into the FA and FI in AI services



\subsection{Failure Analysis in AI Service}

In our detailed exploration of potential failures in AI services, we have identified a broad spectrum of faults. These can be classified into several major categories including data fault, code and software fault, network transmission fault, and external attack. We summarize the types of failures in ~\autoref{tab:fa-service} that can occur in AI services. The \textit{paper} column in ~\autoref{tab:fa-service} shows some representative papers in this study, as do the following tables below.


\textbf{Data Fault.} These faults related to the format, type, and noise of the data can lead to the failure of AI services~\cite{SchelterRBJENGA21,TanCLOnline23,ZhaoCBAConcept22}. For example, incorrect data encoding (e.g., requesting UTF-8 but receiving ASCII) or inappropriate data types (e.g., expecting a string but receiving an integer) can prevent AI services from working properly. The JENGA framework explores the impact of data faults on predictions of machine learning models~\cite{SchelterRBJENGA21}. Furthermore, data drift and concept drift are common problems~\cite{TanCLOnline23}. Data drift refers to a model trained on a specific distribution of data but then encountering a different distribution in practice. However, concept drift occurs as the relationship between features and labels becomes invalid over time. Zhao \textit{et al.}~\cite{ZhaoCBAConcept22} investigate the impact of concept drift on model performance.



\textbf{Development Fault.} The primary reason for service failures is code quality such as bugs, logical faults in code, and poor software design~\cite{LenarduzziSoftware21, GeziciSystematic22,HuCharacterization2021}. Code faults typically originate from coding mistakes. Logical faults in code often involve incorrect algorithm implementation, impacting the accuracy and efficiency of inference. Thus, poor software design affects system performance. Additionally, failures in AI service API calls~\cite{WanAPIs21}, originating from API incompatibility, API change, and API misuse, can lead to AI service failures.


\textbf{Deployment Fault.} During the deployment of AI services, various faults can arise, including outdated models~\cite{OutdatedLLM}, path configuration errors~\cite{Hien2024MLOpsRay}, and inappropriate resource allocation~\cite{Mao23ElasticResource}. These faults can impact system performance and stability. As data change over time, model performance may degrade, necessitating regular updates and retraining to maintain their effectiveness. Path configuration faults can prevent the proper loading of models and data. Inadequate resource allocation~\cite{Mao23ElasticResource}, especially inefficient use of CPU and GPU resources, can lead to decreased system performance and unnecessary waste.

\textbf{Network Transmission Fault.} Failures in network hardware leading to packet delays or losses~\cite{LambertiNetwork2023}. Network congestion occurs when data traffic exceeds the network's bandwidth capacity, preventing data transmission on time. Bandwidth limitations are imposed by the maximum transmission rate of a network connection, often determined by service providers or the capabilities of network hardware. Furthermore, physical damage or configuration faults in network devices can also lead to packet delays or losses.

\textbf{External Attack.} In addition to internal faults within AI services, external attacks can also lead to service failures. These include network attacks such as Distributed Denial of Service (DDoS) attacks~\cite{mattiCybersecurity2023} and Man-in-the-Middle (MITM) attacks~\cite{raotricking2023}, which can not only cause temporary service interruptions but also lead to data leakage or corruption. Moreover, adversarial attacks target the AI model by designing malicious inputs (e.g., meticulously modified images or texts) to deceive the AI into making incorrect actions~\cite{NotAbdelnabi2023,AkhtarMKSAdvances21, QiAdversarial24}. Adversarial attacks are divided into white-box attacks and black-box attacks. In white-box attacks, the deployed model is fully understood, including inputs and architecture, allowing for targeted attacks. In black-box attacks, only the model's inputs and output labels/confidences are accessible.

\begin{table}[t]
\centering
\caption{Failure Analysis in AI Service}
\vspace{-0.1in}
\label{tab:fa-service}
\resizebox{0.9\textwidth}{!}{%
\begin{tabular}{@{}c|c|p{8cm}|c@{}}
\hline
\toprule
\textbf{Group}                                  & \textbf{Failure}                                         & \multicolumn{1}{c|}{\textbf{Description}}                                                                       & \textbf{Paper}                                                                                              \\  \midrule
\multirow{6}{*}{Data}                    & \multirow{1}{*}{Data Quality}                                 & Issues related to the format, type, and noise of data.                                                         & \multirow{1}{*}{\cite{SchelterRBJENGA21}}                                                                    \\  \cline{2-4}
                                                & \multirow{2}{*}{Data Drift}                                   & A model trained on a specific distribution of data but then encountering a different distribution in practice. & \multirow{2}{*}{\cite{TanCLOnline23}}                                                                        \\  \cline{2-4}
                                                & \multirow{2}{*}{Concept Drift}                                & The relationship between features and labels becomes invalid over time.                                        & \multirow{2}{*}{\cite{TanCLOnline23, ZhaoCBAConcept22}}                                                      \\  \midrule
\multirow{2}{*}{Development}           & \multirow{1}{*}{Defective Code}                                 & Logical faults in code, and poor software design.                                                        & \multirow{1}{*}{\cite{LenarduzziSoftware21, GeziciSystematic22,HuCharacterization2021}}                                             \\  \cline{2-4}
                                                & \multirow{1}{*}{Service API Fault}                                  & API Incompatibility, API Change, and API Misuse.                                 & \multirow{1}{*}{\cite{WanAPIs21,CaoUnderstanding2022}}   
                                                \\  \midrule
\multirow{2}{*}{Deployment Fault}  & \multirow{2}{*}{Configuration Fault} & Outdated models, path configuration faults, and inappropriate resource allocation.                                                          & \multirow{2}{*}{\cite{Hien2024MLOpsRay, Mao23ElasticResource,OutdatedLLM}}  \\  \midrule
\multirow{2}{*}{Network Fault}    & \multirow{2}{*}{Network Transmis. Fault}                               & Network congestion, bandwidth limitations, or failures in network hardware.                                    & \multirow{2}{*}{\cite{LambertiNetwork2023}}                                                                  \\  \midrule
\multirow{4}{*}{External Attack}         & \multirow{2}{*}{Network Attack}                               & Lead to temporary service interruptions and data leakage or corruption.                                        & \multirow{2}{*}{\cite{mattiCybersecurity2023, raotricking2023}}                                              \\  \cline{2-4}
                                                & \multirow{2}{*}{Adversarial Attack}                           & Deceive the AI into making incorrect actions through malicious inputs.                                         & \multirow{2}{*}{\cite{NotAbdelnabi2023,AkhtarMKSAdvances21, QiAdversarial24}}                                                 \\  \bottomrule
\end{tabular}%
}
\end{table}

\begin{table}[t]
\centering
\caption{Fault Injection in AI Service}
\vspace{-0.1in}
\label{tab:fi-service}
\resizebox{\textwidth}{!}{%
\begin{tabular}{@{}c|c|p{7cm}|c}
\toprule
\textbf{Group}                  & \textbf{Fault}                & \multicolumn{1}{c|}{\textbf{Description}}                                                                                               & \textbf{Tools or Methods} \\    \midrule
\multirow{4}{*}{Data}  & \multirow{2}{*}{Data Perturbation}       & Introduce noise or modifying data artificially can simulate uncertainties in real-world data.                      & \multirow{2}{*}{\begin{tabular}[c]{@{}c@{}}NumPy~\cite{numpyGithub2024}, Scikit-learn~\cite{scikit-learnGithub2024}, \\ JENGA~\cite{SchelterRBJENGA21}\end{tabular}  }  \\ \cline{2-4}
                       & \multirow{2}{*}{Data and Concept Drift}  & Simulate sudden, gradual, and incremental drifts in data streams.                                                  & \multirow{2}{*}{\begin{tabular}[c]{@{}c@{}} Scikit-multiflow~\cite{Jacobskmultiflow2018}, \\ MOA~\cite{BifetMOA2010} \end{tabular} }  \\    \midrule
\multirow{2}{*}{Service API}   & \multirow{2}{*}{Service API Fault}     &  Leverage Envoy to introduce API return errors into service communication.                                                & \multirow{2}{*}{Istio~\cite{IstioGithub2024}, MicroFI~\cite{MicroFI2024TDSC}} \\ \midrule                       
\multirow{2}{*}{Network}   & \multirow{2}{*}{Low-Quality Network}     & Simulate network delays, jitter, packet loss, and reordering.                                                      & \multirow{2}{*}{Toxiproxy~\cite{ToxiproxyGithub2024}, ChaosBlade~\cite{chaosbladeGithub2024}} \\ \midrule
\multirow{4}{*}{\begin{tabular}[c]{@{}c@{}}External \\ Attack \end{tabular}}   & \multirow{2}{*}{Adversarial Attack}      & Use known information and relevant patterns to attack the model.                                                   & \multirow{2}{*}{\begin{tabular}[c]{@{}c@{}}FGSM~\cite{GoodfellowSSFGSM14}, PGD~\cite{MadryMSTVPGD18}, \\ DI-AA~\cite{WangLCRWDIAA22}, Grey-box attack~\cite{RazGrayBox23}\end{tabular} }              \\   
                       \cline{2-4}
                       & \multirow{2}{*}{Prompt Attack}           & Guide models to generate special outputs by adding prompts to the input text.                                      & \multirow{2}{*}{\begin{tabular}[c]{@{}c@{}} IPI~\cite{NotAbdelnabi2023}, PromptAid~\cite{mishrapromptaid2023}, \\  HouYi~\cite{liuprompt2023}, Goal-guided attack~\cite{zhanggoal2024}\end{tabular} } \\ \bottomrule

\end{tabular}%
}
\vspace{-0.2in}
\end{table}

\subsection{Fault Injection in AI Service}
Fault injection in AI service layer encompasses four dimensions, including data, API service, network transmission, and external attack. By simulating diverse fault scenarios in these dimensions, it is possible to assess the system's robustness and reliability, thereby preventing inaccurate predictions or even system failures. ~\autoref{tab:fi-service} illustrates the current FI tools in the AI service layer, followed by a description of each of them.

\textbf{Data Fault.} Data perturbation is the most intuitive method of FI at the data dimension. Introducing noise or modifying the data manually can simulate uncertainties in real-world data. It can be achieved using tools such as NumPy~\cite{numpyGithub2024}, Scikit-learn~\cite{scikit-learnGithub2024}, and so on. JENGA~\cite{SchelterRBJENGA21} is a framework that studies the impact of data faults (e.g., missing values, outliers, typing errors, and noisy) on the predictions of machine learning models. Furthermore, data and concept drift can be simulated through carefully designed data disturbances. Moreover, tools such as scikit-multiflow~\cite{Jacobskmultiflow2018} and MOA~\cite{BifetMOA2010} enable the simulation of sudden, gradual, and incremental drifts in data streams. 

\textbf{Service API Fault.}  Istio, an open-source service mesh, addresses provides robust traffic management features. Istio's fault injection capabilities are primarily exposed through its Service API, which allows users to define fault injection rules declaratively~\cite{IstioGithub2024,MicroFI2024TDSC}. These rules are specified within Istio's VirtualService resources, which are then propagated to the Envoy proxies deployed as sidecars alongside each service instance. However, deploying and managing Istio can add complexity to the AI infrastructure. The learning curve for Istio is steep, and managing its components alongside AI services can be challenging.

\textbf{Network Fault.} Common network fault injections include network delays, network jitter, packet loss, and reordering. The injection of these network faults may result in the delayed or non-response of user requests for AI services. Toxiproxy~\cite{ToxiproxyGithub2024} is a TCP proxy used to simulate network and system conditions for chaos and resilience testing. ChaosBlade~\cite{chaosbladeGithub2024} is an  open source experimental injection tool that adheres to the principles of chaos engineering and chaos experimental models, supporting a rich array of experimental scenarios. Istio can also leverage the Envoy's advanced traffic management capabilities to simulate network conditions such as delay, packet loss, and service unavailability~\cite{IstioFault2024}.

\textbf{External Attack.} Adversarial attacks are deliberate attacks on the AI service, including both white-box and black-box attacks. White-box attacks leverage model structure and parameter information to strategically generate adversarial samples, such as FGSM~\cite{GoodfellowSSFGSM14}, PGD~\cite{MadryMSTVPGD18}, and DI-AA~\cite{WangLCRWDIAA22}. Black-box attacks lack insight into the model and rely on observing model inputs and outputs to create adversarial samples, such as HopSkipJump~\cite{ChenJWHopSkipJumpAttack20}, PopSkipJump~\cite{SimonPopSkipJump21}, and GeoDA~\cite{RahmatiMFDGeoDA20}. Grey-box attacks utilize partial information to generate adversarial samples. Raz \textit{et al.}~\cite{RazGrayBox23} attack image-to-text models based on adversarial perturbations. In addition to traditional methods, prompt fault injection has become popular for large models. It guides models to generate special outputs by adding prompts to the input text. Greshake \textit{et al.}~\cite{NotAbdelnabi2023} use indirect prompt injection to exploit LLM-integrated applications and systematically investigate impacts and vulnerabilities, including data theft, worming, information ecosystem contamination, and other novel security risks. PromptAid~\cite{mishrapromptaid2023} can introduce keyword and paraphrasing perturbations into prompts. HouYi~\cite{liuprompt2023} conducts prompt fault injections in a black-box manner. Zhang \textit{et al.}~\cite{zhanggoal2024} propose a goal-guided generative prompt injection attack on LLMs.


\subsection{Gap between Failure Analysis and Fault Injection in AI Service}

\begin{table}[t]
\caption{Gap between Failure Analysis and Fault Injection in AI Service}
\centering
\vspace{-0.1in}
\label{tab:Service-gap}
\resizebox{0.9\textwidth}{!}{
\begin{tabular}{c|c|c|c|c|c|c|c|c@{}}
\toprule
Group & Failure                    & JENGA  & MOA    & FGSM   & PromptAid  & Istio  & ChaosBlade &  Covered  \\  \midrule
\multirow{3}{*}{Data} & Data Quality               & \textbf{\CheckmarkBold}&        &        &             &        & & True            \\ \cline{2-9}
& Data Drift                 &        & \textbf{\CheckmarkBold}&        &                     &        &   & True         \\ \cline{2-9}
& Concept Drift              &        & \textbf{\CheckmarkBold}&        &                     &        & & True            \\ \midrule
\multirow{3}{*}{Development}  & Defective Code &        &        &        &                      &        &      & False       \\ \cline{2-9}
& Configuration Fault          &        &        &        &                      &    &      &   False     \\ \cline{2-9}
& Service API Fault          &        &        &        &                      &  \textbf{\CheckmarkBold}     &      & True       \\ \midrule
Network & Network Transmission Fault &        &        &        &                      &  \textbf{\CheckmarkBold}     & \textbf{\CheckmarkBold}& True     \\ \midrule
\multirow{2}{*}{Attack} & Network Attack             &        &        &        &                      &        & \textbf{\CheckmarkBold} & True    \\ \cline{2-9}
 & Adversarial Attack         &        &        & \textbf{\CheckmarkBold}& \textbf{\CheckmarkBold}            &        &          & True   \\ \bottomrule
\end{tabular}%
}
\vspace{-0.1in}
\end{table}

Based on the comparative analysis presented in Table~\ref{tab:Service-gap}, which outlines the capabilities of various FI tools in addressing diverse failure modalities within AI services, several critical insights regarding the discrepancies between FA and FI can be elucidated. 

\textbf{Incomplete Coverage.} The scrutinized FI tools, in aggregate, encompass a substantial proportion of the delineated fault types. Nonetheless, certain fault types (e.g., ``Defective Code'' ) remain unaddressed by any of the existing tools. This revelation underscores the exigency for subsequent research to ameliorate these deficiencies in fault simulation.

\textbf{Diversity of Tools.} Each FI tool exhibits a predilection for addressing specific fault types. For instance, JENGA is attuned to data quality faults, whereas MOA is adept at handling data and concept drift faults. This diversity implies that practitioners may be compelled to deploy a suite of tools to achieve a comprehensive simulation and analysis of failures within AI services, contingent upon the specific fault types of interest.

\textbf{New Emerging Fault Types.} The incorporation of fault types such as ``Prompt Injection Attacks'' accentuates the burgeoning significance of accounting for external factors and security facets in the FA of AI services. 
Nevertheless, traditional FI tools (e.g., ChaosBlade) are designed for cloud computing services and do not consider the specific scenarios of AI systems. As AI services become increasingly interconnected and susceptible to a myriad of threats, it is imperative to cultivate FI tools capable of effectively simulating failures emanating from these nascent domains.

\section{Failure Analysis and Fault Injection in AI Model}\label{sec:model}
AI model layer is a crucial component in AI systems, residing beneath the AI service layer. This layer is responsible for managing the various AI models and algorithms that power the AI capabilities of the system. Similar to the AI service layer, potential failures at this layer can include unavailability of the model or incorrect outputs. In this section, we delve into the FA and FI in AI models. 


\subsection{Failure Analysis in AI Model}
Recent research has explored multiple reasons for AI model failures. Researchers analyze various sources, including GitHub commits, Stack Overflow posts, and expert interviews. These analyses have provided crucial insights into enhancing the reliability and robustness of AI systems. For example, Humbatova \textit{et al.}~\cite{HumbatovaTaxonomy2020} categorize faults within deep learning systems by examining 1,059 GitHub commits and issues of AI systems, while Islam \textit{et al.}~\cite{IslamComprehensive2019} analyze error-prone stages, bug types, and root causes in deep learning pipelines from 2,716 Stack Overflow posts and 500 GitHub bug fix commits. Additionally, Nikanjam \textit{et al.}~\cite{Nikanjam2022reinforcement} classify faults in deep reinforcement learning programs from 761 documents. We focus on the training and testing phases of AI models, where faults are categorized as data faults, model hyperparameter faults, and model structure and algorithm faults, as shown in Table~\ref{tab:fa-model}.

\begin{table}[t]
\centering
\caption{Failure Analysis in AI Model}
\label{tab:fa-model}
\vspace{-0.1in}
\resizebox{\textwidth}{!}{%
\begin{tabular}{@{}c|c|p{8cm}|c@{}}
\hline
\toprule
\textbf{Group}                                        & \textbf{Failure}                               & \multicolumn{1}{c|}{\textbf{Description}}                                                                      & \textbf{Paper} \\    \midrule
\multirow{3}{*}{\begin{tabular}[c]{@{}c@{}}Data\end{tabular}}                 & \multirow{1}{*}{Data Quality}                         & Low-quality data leads to poor model performance.                                                & \multirow{1}{*}{\cite{CroftData23, WhangRSLData23, LiangCorrection22}}                \\   \cline{2-4}
                                                      & \multirow{2}{*}{Data Preprocessing Fault}           & The inadequate handling of data noise, damage, loss, and inconsistency.                                                      & \multirow{2}{*}{\cite{CaoUnderstanding2022,Saikatpreprocessing2022, maharanareview2022}}                \\   \midrule
\multirow{4}{*}{\begin{tabular}[c]{@{}c@{}}Model \\ Hyperparameter\end{tabular}}          & \multirow{2}{*}{\begin{tabular}[c]{@{}c@{}}Inappropriate Layer and \\ Neuron Quantity\end{tabular}}     & Incorrectly setting the number of layers and neurons can affect the model's parameter count and performance.                                                 & \multirow{2}{*}{\cite{CaoUnderstanding2022,Bashaconnected20, uzaireffects2020, MenShortGPT2024, GurneeNeurons2023}}                \\   \cline{2-4}
                                                      & \multirow{2}{*}{\begin{tabular}[c]{@{}c@{}}Inappropriate Learning Rate,  \\  Epochs, and Batch Size\end{tabular}} & Influence the training speed and model performance (overfitting and underfitting).        & \multirow{2}{*}{\cite{HeBatch19, shafiexploring2023}}                \\   \midrule
\multirow{6}{*}{\begin{tabular}[c]{@{}c@{}}Model \\ Structure and \\   Algorithm \end{tabular} } & \multirow{1}{*}{Misuse Neural Networks}             & Inappropriate types of neural networks.                     &    \multirow{1}{*}{\cite{ShiriOverview2023}}                \\  \cline{2-4}
                                                      & \multirow{1}{*}{Misuse Activation Function}                  & Introduce non-linearity to enhance model fitting ability.                                                                  & \multirow{1}{*}{\cite{DubeyActivation22}}                \\   \cline{2-4}
                                                      & \multirow{1}{*}{Misuse Regularization}                       & Inappropriate regularization lead to overfitting.                                                                        & \multirow{1}{*}{\cite{Tianregularization22}}                \\   \cline{2-4}
                                                      & \multirow{1}{*}{Misuse Optimizer}                            & Influence the training speed and model performance.                                       & \multirow{1}{*}{\cite{hajicomparison2021}}                \\   \cline{2-4}
                                                      & \multirow{1}{*}{Misuse Loss Function}                        & Affect the speed and degree of convergence in training.                     & \multirow{1}{*}{\cite{wangloss2020}}                \\   \cline{2-4}
                                                      & \multirow{1}{*}{Dataset Partitioning Fault}                 & Insufficient data for training and validation. & \multirow{1}{*}{\cite{murainaideal2022}}                 \\ \bottomrule
\end{tabular}%
}
\vspace{-0.2in}
\end{table}

\textbf{Data Fault.} The quality of training data is pivotal for successful model training~\cite{yin2023data}. Alice \textit{et al.}~\cite{CroftData23} conduct a systematic analysis of data quality attributes (accuracy, uniqueness, consistency, completeness, and timeliness) across five software bug datasets and find that 20-71\% of data labels are inaccurate, which could severely hinder model training in extreme cases. Some studies~\cite{WhangRSLData23, LiangCorrection22} have examined challenges faced by data quality. Furthermore, data preprocessing and augmentation are crucial.
Raw data are susceptible to noise, damage, loss, and inconsistency, thus necessitating preprocessing steps (e.g., data cleaning, integration, transformation, and reduction) to facilitate easier knowledge extraction from datasets. 
Data augmentation aims to expand the training dataset through specific transformations to enhance the model's generalizability. Das~\cite{Saikatpreprocessing2022} lists ten common faults in data preprocessing, while Maharana \textit{et al.}~\cite{maharanareview2022} discuss various data preprocessing and data augmentation techniques to enhance model performance.


\textbf{Model Hyperparameter Fault.} 
The parameters of an AI model include both pre-training hyperparameters (e.g., the number of hidden layers, batch size, and learning rate) and post-training model parameters (e.g., network weights and biases). 
Basha \textit{et al.}~\cite{Bashaconnected20} examine the effects of different numbers of fully connected layers on convolutional neural networks (CNNs). Uzair \textit{et al.}~\cite{uzaireffects2020} investigate how the number of hidden layers affects the efficiency of neural networks. ShortGPT~\cite{MenShortGPT2024} points out that many layers in LLMs exhibit high redundancy, with some layers playing minimal roles. Gurnee \textit{et al.}~\cite{GurneeNeurons2023} utilize seven different models (ranging from 70 million to 6.9 billion parameters) to study the sparsity of activations in LLM neurons. Additionally, the learning rate (LR), epochs, and batch size (BS) influence the training speed and the performance of the trained model. He \textit{et al.}~\cite{HeBatch19} indicate that the ratio of batch size to learning rate should not be too large to ensure good generalization. Shafi \textit{et al.}~\cite{shafiexploring2023} explore the optimization of hyperparameters, including the learning rate, batch size, and epochs, as well as their interrelationships.

\textbf{Model Structure and Algorithm Fault.} 
Shiri \textit{et al.}~\cite{ShiriOverview2023} investigate various aspects of different models and evaluate their performance on three public datasets. Activation functions are crucial for introducing non-linearity. Dubey \textit{et al.}~\cite{DubeyActivation22} compare the performance of 18 distinct activation functions (e.g., Sigmoid, Tanh, and ReLU) on various datasets. Model training requires regularization to avoid overfitting. Tian \textit{et al.}~\cite{Tianregularization22} compare different regularization techniques, including sparse regularization, low-rank regularization, dropout, batch normalization, and others. They discuss the selection of regularization techniques for specific tasks. The selection of optimizers significantly affects model performance and training speed. Haji \textit{et al.}~\cite{hajicomparison2021} compare various optimizers such as SGD, Adam, AdaGrad, etc. They highlight the advantages and disadvantages of these optimizers in terms of training speed, convergence rate, and performance. The loss function is also essential for minimizing the discrepancy between predicted results and target values. Wang \textit{et al.}~\cite{wangloss2020} introduce 31 loss functions from five aspects: classification, regression, unsupervised learning of traditional machine learning, object detection and face recognition of deep learning. Furthermore, the ratio of training to validation data must be considered to ensure that the model has enough data for learning while the validation data is adequate for model adjustments~\cite{murainaideal2022}.

\begin{table}[t]
\centering
\caption{Fault Injection in AI Model}
\vspace{-0.1in}
\label{tab:fi-model}
\resizebox{0.9\textwidth}{!}{%
\begin{tabular}{@{}c|l|c|c|c|c@{}}
\hline
\toprule
\textbf{Group}                         & \multicolumn{1}{c|}{ \textbf{Fault}} & \textbf{DeepMutation}~\cite{LeiDeepMutation2018} & \textbf{DeepMutation++}~\cite{Hu0XY0ZDeepMutation19} & \textbf{MuNN}~\cite{ShenMuNN2018} & \textbf{DeepCrime}~\cite{HumbatovaDeepCrime2021} \\ \midrule
\multirow{7}{*}{Data}         & Duplicates training data                    & \textbf{\CheckmarkBold}                 &                         &               &                   \\  \cline{2-6}
                                       & Shuffle training data                       & \textbf{\CheckmarkBold}                 &                         &               &                    \\ \cline{2-6}
                                       & Change labels of training data              & \textbf{\CheckmarkBold}                 &                         &               & \textbf{\CheckmarkBold}              \\ \cline{2-6}
                                       & Remove part of training data             & \textbf{\CheckmarkBold}                 &                         &               & \textbf{\CheckmarkBold}              \\ \cline{2-6}
                                       & Unbalance training data                     &                       &                         &               & \textbf{\CheckmarkBold}              \\ \cline{2-6}
                                       & Add noise to training data                  & \textbf{\CheckmarkBold}                 &                         &               & \textbf{\CheckmarkBold}              \\ \cline{2-6}
                                       & Make output classes overlap                 &                       &                         &               & \textbf{\CheckmarkBold}              \\  \cline{1-6}
\multirow{4}{*}{Hyperparameters}       & Change batch size                           &                       &                         &               & \textbf{\CheckmarkBold}              \\  \cline{2-6}
                                       & Decrease learning rate                      &                       &                         &               & \textbf{\CheckmarkBold}              \\  \cline{2-6}
                                       & Change number of epochs                     &                       &                         &               & \textbf{\CheckmarkBold}              \\  \cline{2-6}
                                       & Disable data batching                       &                       &                         &               & \textbf{\CheckmarkBold}              \\   \cline{1-6}
\multirow{3}{*}{\begin{tabular}[c]{@{}c@{}}Activation \\ Function\end{tabular}}   & Change activation function                  &                       &                         & \textbf{\CheckmarkBold}         & \textbf{\CheckmarkBold}              \\  \cline{2-6}
                                       & Remove activation function                  & \textbf{\CheckmarkBold}                 &                         &               & \textbf{\CheckmarkBold}              \\  \cline{2-6}
                                       & Add activation function            &                       &                         &               & \textbf{\CheckmarkBold}              \\   \cline{1-6}
\multirow{5}{*}{Regularisation}        & Add weights regularisation                  &                       &                         &               & \textbf{\CheckmarkBold}              \\  \cline{2-6}
                                       & Change weights regularisation               &                       &                         &               & \textbf{\CheckmarkBold}              \\  \cline{2-6}
                                       & Remove weights regularisation               &                       &                         &               & \textbf{\CheckmarkBold}              \\  \cline{2-6}
                                       & Change dropout rate                         &                       &                         &               & \textbf{\CheckmarkBold}              \\  \cline{2-6}
                                       & Change patience parameter                   &                       &                         &               & \textbf{\CheckmarkBold}              \\   \cline{1-6}
\multirow{5}{*}{Weights}               & Change weights initialisation               &                       &                         &               & \textbf{\CheckmarkBold}              \\ \cline{2-6}
                                       & Add bias to a layer                         &                       &                         & \textbf{\CheckmarkBold}         & \textbf{\CheckmarkBold}              \\  \cline{2-6}
                                       & Remove bias from a layer                    &                       &                         &               & \textbf{\CheckmarkBold}              \\  \cline{2-6}
                                       & Change weights                              & \textbf{\CheckmarkBold}                     & \textbf{\CheckmarkBold}                   & \textbf{\CheckmarkBold}         &                    \\ \cline{2-6}
                                       & Shuffle weights                             & \textbf{\CheckmarkBold}                     & \textbf{\CheckmarkBold}                   &               &                    \\   \cline{1-6}
Loss function                          & Change loss function                        &                       &                         &               & \textbf{\CheckmarkBold}              \\   \cline{1-6}
\multirow{2}{*}{\begin{tabular}[c]{@{}c@{}}Optimisation \\ Function\end{tabular} } & Change optimisation function                &                       &                         &               & \textbf{\CheckmarkBold}              \\ \cline{2-6}
                                       & Change gradient clipping                    &                       &                         &               & \textbf{\CheckmarkBold}              \\   \cline{1-6}
Validation                             & Remove validation set                       &                       &                         &               & \textbf{\CheckmarkBold}              \\   \cline{1-6}
\multirow{3}{*}{Layers}                & Remove layer                                & \textbf{\CheckmarkBold}                 & \textbf{\CheckmarkBold}                   &               &                    \\ \cline{2-6}
                                       & Add layer                                   & \textbf{\CheckmarkBold}                 & \textbf{\CheckmarkBold}                   &               &                    \\ \cline{2-6}
                                       & Duplicate one layer                         &                       & \textbf{\CheckmarkBold}                   &               &                    \\   \cline{1-6}
\multirow{5}{*}{Neuron}                & Delete Input Neuron                         &                       &                         & \textbf{\CheckmarkBold}         &                    \\ \cline{2-6}
                                       & Delete Hidden Neuron                        &                       &                         & \textbf{\CheckmarkBold}         &                    \\ \cline{2-6}
                                       & Block a neuron effect 0                     &   \textbf{\CheckmarkBold}                   & \textbf{\CheckmarkBold}                   &               &                    \\ \cline{2-6}
                                       & Invert the activation status                &  \textbf{\CheckmarkBold}                     & \textbf{\CheckmarkBold}                   &               &                    \\ \cline{2-6}
                                       & Switch two neurons        &  \textbf{\CheckmarkBold}                    & \textbf{\CheckmarkBold}                   &               &                    \\   \cline{1-6}
\multirow{9}{*}{RNN Specific}          & Fuzz weights                                &                       & \textbf{\CheckmarkBold}                   &               &                    \\ \cline{2-6}
                                       & Reduce weight's precision                   &                       & \textbf{\CheckmarkBold}                   &               &                    \\ \cline{2-6}
                                       & Clear the state to 0                        &                       & \textbf{\CheckmarkBold}                   &               &                    \\ \cline{2-6}
                                       & Reset state to previous state               &                       & \textbf{\CheckmarkBold}                   &               &                    \\ \cline{2-6}
                                       & Fuzz state value                            &                       & \textbf{\CheckmarkBold}                   &               &                    \\ \cline{2-6}
                                       & Reduce state value's precision              &                       & \textbf{\CheckmarkBold}                   &               &                    \\ \cline{2-6}
                                       & Clear the gate value to 0                   &                       & \textbf{\CheckmarkBold}                   &               &                    \\ \cline{2-6}
                                       & Fuzz gate value                             &                       & \textbf{\CheckmarkBold}                   &               &                    \\ \cline{2-6}
                                       & Reduce gate value's precision               &                       & \textbf{\CheckmarkBold}                   &               &                    \\   \cline{1-6}
\multicolumn{2}{c|}{ \textbf{Number}  }                                                & 13                     & 17                      & 5             & 24                    \\ \bottomrule     
\end{tabular}%
}
\vspace{-0.1in}
\end{table}

\subsection{Fault Injection in AI Model}


Fault injection in AI models typically involves interfering with the training process to create models with inferior performance. This technique was called mutation testing. The concept of mutation testing that we will discuss next is analogous to fault injection. A common method for conducting mutation testing on AI models involves designing mutation operators that introduce faults into the training data or the model training program, and then analyzing the behavioral differences between the original and mutated models.

Recent studies in mutation testing have made significant contributions. We have summarized these works as shown in~\autoref{tab:fi-model}. DeepMutation~\cite{LeiDeepMutation2018} designs 13 mutation operators that inject faults into both the training data and code of deep learning models. DeepMutation++~\cite{Hu0XY0ZDeepMutation19} combines DeepMutation (eight model-level operators for FNN models) and proposes nine new operators specifically for RNN models, enabling both static mutations in FNNs and RNNs and dynamic mutations in RNNs. MuNN~\cite{ShenMuNN2018} develops five mutation operators, focusing on model-level fault injection. DeepCrime~\cite{HumbatovaDeepCrime2021} implements 24 deep learning mutation operators to test the robustness of deep learning systems, including training data operators, hyperparameters operators, activation function operators, regularization operators, weight operators, loss function operators, optimization operators, and validation operators. Additionally, some studies have focused on mutation testing in reinforcement learning~\cite{TambonMNKAMutation23, LuSSmutation22} and unsupervised learning~\cite{LuSSSMTUL22}.

\subsection{Gap between Failure Analysis and Fault Injection in AI Model}
\begin{table}[t]
\caption{Gap between Failure Analysis and Fault Injection in AI Model}
\centering
\vspace{-0.1in}
\label{tab:Model-gap}
\resizebox{1\textwidth}{!}{%
\begin{tabular}{c|c|c|c|c|c|c@{}}
\toprule
Group & Failure                                             & DeepMutation & DeepMutation++ & MuNN & DeepCrime & Covered\\   \midrule
\multirow{2}{*}{Data} & Data Quality                                        & \textbf{\CheckmarkBold}        &                &      &  \textbf{\CheckmarkBold} & True    \\ \cline{2-7}
& Data Preprocessing Fault                            & \textbf{\CheckmarkBold}        &                &      & \textbf{\CheckmarkBold} & True     \\ \midrule
\multirow{2}{*}{\begin{tabular}[c]{@{}c@{}}Model \\ Hyperparameter\end{tabular}} & Layer and Neuron Quantity Fault                     & \textbf{\CheckmarkBold}        & \textbf{\CheckmarkBold}          & \textbf{\CheckmarkBold}&  & True          \\ \cline{2-7}
& Inappropriate LR, Epochs, and BS &              &                &      & \textbf{\CheckmarkBold}    & True   \\ \midrule
\multirow{6}{*}{\begin{tabular}[c]{@{}c@{}}Model \\ Structure and \\   Algorithm \end{tabular} } & Misuse Neural Networks                              &  \textbf{\CheckmarkBold}            &    \textbf{\CheckmarkBold}            &  \textbf{\CheckmarkBold}    &   & True        \\ \cline{2-7}
& Misuse Activation Function                          & \textbf{\CheckmarkBold}        &                & \textbf{\CheckmarkBold}& \textbf{\CheckmarkBold}     & True   \\ \cline{2-7}
& Misuse Regularization                               &              &                &      & \textbf{\CheckmarkBold}     & True   \\ \cline{2-7}
& Misuse Optimizer                                    &              &                &      & \textbf{\CheckmarkBold}     & True   \\ \cline{2-7}
& Misuse Loss Function                                &              &                &      & \textbf{\CheckmarkBold}     & True   \\ \cline{2-7}
& Dataset Partitioning Fault                          &              &                &      & \textbf{\CheckmarkBold}   & True     \\ \bottomrule
\end{tabular}%
}
\vspace{-0.1in}
\end{table}

Based on the comparative analysis presented in ~\autoref{tab:Model-gap}, which delineates the capabilities of various FI tools in addressing diverse failure modalities within AI models, two critical insights regarding the discrepancies between FA and FI can be elucidated:

\textbf{Differentiated Focus.} The distinct FI tools appear to focus on disparate facets of AI model failures. For instance, DeepMutation and DeepCrime are adept at handling data quality and preprocessing faults, whereas MuNN is tailored towards layer and neuron quantity faults. This specialization implies that the selection of an FI tool should be contingent upon the specific fault types targeted for analysis.

\textbf{Coverage Inconsistency.} ~\autoref{tab:Model-gap} reveals a disparity in coverage, with certain fault types, such as ``Layer and Neuron Quantity Fault'' and ``Misuse Activation Function'', addressed by multiple tools, while others, like ``Inappropriate LR, Epochs, and BS'', are solely within the purview of DeepCrime. This inconsistency may reflect the inherent challenges associated with simulating specific fault types or indicate a relative lack of focus within the FI.





\section{Failure Analysis and Fault Injection for AI Framework}\label{sec:framework}
AI framework layer, including TensorFlow~\cite{MartTensorFlow2016}, PyTorch~\cite{AdamPyTorch2019}, and Keras~\cite{keras},  acts as a bridge between the AI model layer and the underlying hardware and system infrastructure.
Akin to other software systems, these AI frameworks are susceptible to a variety of faults~\cite{FlorianSilent2024, ZengyangUnderstanding2023}. 
Failures at the AI framework layer can lead to unavailability, incorrect output, and poor performance perceived by users.
Thus, ensuring the robustness of AI frameworks is crucial for the reliability of AI systems. This section is dedicated to analyzing faults in the AI framework layer.


\subsection{Failure Analysis in AI Framework}
Over recent years, a significant volume of research~\cite{ZhangAn2018,MartTensorFlow2016,IslamComprehensive2019,HumbatovaTaxonomy2020,ChenToward2023,FlorianSilent2024,ZengyangUnderstanding2023,Yangcomprehensive2022} has been dedicated to the analysis of failures in AI frameworks. Research on AI failure can be primarily bifurcated into two categories, including failures arising from the usage of AI frameworks and failures stemming from the frameworks' implementation.
We summarize the types of failures in Table~\ref{tab:fa-framework} that can occur in both the utilization and the implementation of AI frameworks.


\begin{table}[t]
\centering
\caption{Failure analysis in AI Framework}
\vspace{-0.1in}
\resizebox{\textwidth}{!}{%
\begin{tabular}{@{}c|l|p{9cm}|c@{}}
\hline
\toprule
\textbf{Group}                                              & \multicolumn{1}{c|}{\textbf{Failure}}                           & \multicolumn{1}{c|}{\textbf{Description}}                                                         & \textbf{Paper}                                                           \\  \midrule
\multirow{3}{*}{\begin{tabular}[c]{@{}c@{}}Data\end{tabular} }                  & \multirow{1}{*}{Tensor Alignment Fault}                     & Tensors do not align as expected, leading to shape mismatches.                      & \multirow{1}{*}{\cite{ZhangAn2018,HumbatovaTaxonomy2020}}                        \\\cline{2-4} 
                                                   & \multirow{1}{*}{Input Format Fault}               & The shape or type of input data mismatches the expected format.         & \multirow{1}{*}{\cite{HumbatovaTaxonomy2020,IslamComprehensive2019,Wang2022Numerical}}             \\\cline{2-4}
                                                   & \multirow{1}{*}{Dataloader Crash Fault}               & Dataloader crashes due to memory leak in multipleworker.         & \multirow{1}{*}{\cite{Hu2024Characterization}}             \\  \midrule
                                                   
\multirow{3}{*}{API}                         & \multirow{1}{*}{API Usage Fault}                    & API is used in a way that does not conform to the correct logic.               & \multirow{1}{*}{\cite{HumbatovaTaxonomy2020,Wang2022Numerical}}                                    \\\cline{2-4} 
                                                   & \multirow{1}{*}{API Compatibility Fault}              & Different APIs are not compatible with each other.                  & \multirow{1}{*}{\cite{ZhangAn2018,IslamComprehensive2019,HumbatovaTaxonomy2020}} \\\cline{2-4} 
                                                   & \multirow{1}{*}{API Version Fault}                   &  API version is incompatible with the code or dependencies.       & \multirow{1}{*}{\cite{ZhangAn2018,IslamComprehensive2019,HumbatovaTaxonomy2020}} \\  \midrule
\multirow{4}{*}{\begin{tabular}[c]{@{}c@{}}Configuration\end{tabular}  }                   & \multirow{1}{*}{Framework Config. Fault }   & Incorrect configuration when using a framework.                    & \multirow{1}{*}{\cite{ZhangAn2018, Yangcomprehensive2022, Quan2022Towards,Du2020TensorFlow}}                       \\\cline{2-4} 
                                                   & \multirow{1}{*}{Device Config. Fault}     & Inability to leverage computing devices for optimal performance.   & \multirow{1}{*}{\cite{HumbatovaTaxonomy2020, Yangcomprehensive2022}}             \\\cline{2-4}  
& \multirow{2}{*}{Environment Config. Fault}     & Environmental configuration faults when developing and deploying an AI framework.   & \multirow{2}{*}{\cite{ZengyangUnderstanding2023, ZhangEmpirical2020, ChenToward2023,Du2020TensorFlow}}             \\                                                     
                                                   \midrule
\multirow{4}{*}{\begin{tabular}[c]{@{}c@{}}Performance \end{tabular}  }                 & \multirow{2}{*}{Memory Management Fault}                & Faults occur when managing memory between heterogeneous devices.     & \multirow{2}{*}{\cite{TarekAn2022,Yangcomprehensive2022, Quan2022Towards,liu2022ISSRE}}                        \\\cline{2-4} 
                                                   & \multirow{1}{*}{Parallelism Fault}               & Includes insufficient parallelism and excessive parallelism.         & \multirow{1}{*}{\cite{TarekAn2022}}                                              \\\cline{2-4} 
                                                   & \multirow{1}{*}{Operator Inefficiency Fault}          & Trade-off of using different linear algebra operators.    & \multirow{1}{*}{\cite{PyTorch42265}}                                             \\  
                                                   \midrule
\multirow{2}{*}{Code}                        & \multirow{1}{*}{Syntax Fault}                          & Faults occur when implementing and using AI framework. & \multirow{1}{*}{\cite{ChenToward2023, FlorianSilent2024} }                       \\\cline{2-4} 
                                                   & \multirow{1}{*}{Cross-Language Fault}     & Faults occur when utilizing multi-programming-language.               & \multirow{1}{*}{\cite{ZengyangUnderstanding2023}}                            \\  
                                                   
                                                   \midrule
\multirow{2}{*}{\begin{tabular}[c]{@{}c@{}}Algorithm\end{tabular} }                   &  \multirow{1}{*}{Implementation Logic Fault}      & Faults present in the algorithm's implementation.                     & \multirow{1}{*}{\cite{ChenToward2023, FlorianSilent2024, Quan2022Towards}} 
\\
\cline{2-4} 
                                                   & \multirow{1}{*}{Algorithm Inefficiency Fault} & Algorithms implemented using outdated or inefficient methods.                    & \multirow{1}{*}{\cite{TarekAn2022,JiaTensorFlow2020}}      \\ 
                                                   
\bottomrule
\end{tabular}%
}
\label{tab:fa-framework}
\vspace{-0.1in}
\end{table}

\textbf{Data Fault} may occur during the data input stage of an AI model. This type of fault is typically caused by unaligned tensors or incorrect formatting of input data. For example, it could occur when inputting a tensor array instead of an individual element from the array, or when mistakenly using a transposed tensor instead of the original tensor~\cite{ZhangAn2018,HumbatovaTaxonomy2020}. Even more critically, the shape or type of the data inputted to the model may completely mismatch the expected input by the model which leads to the model unable to run correctly~\cite{HumbatovaTaxonomy2020,IslamComprehensive2019}. One additional fault that occurs during the data input stage is the data loader crash fault~\cite{Hu2024Characterization}. This fault primarily occurs in training tasks of LLM in multiple workers. This issue arises from a gradual memory leak due to PyTorch's implementation of data loader, which is caused by the copy-on-write mechanism used in the multiprocessing fork start method, combined with a suboptimal design of the Python list.

\textbf{API Fault} typically occurs during the call to APIs provided by AI frameworks. Such faults may due to the using of an API in a way that does not conform to the logic set out by developers of the framework~\cite{HumbatovaTaxonomy2020}. Indeed, lack of inter-API compatibility and versioning issues could be one of the main culprits~\cite{ZhangAn2018,IslamComprehensive2019,HumbatovaTaxonomy2020}. When different APIs are not compatible with each other or when the version of the API being used is not compatible with the requirements of the code or dependencies, it can result in API faults.

\textbf{Configuration Fault} typically occurs due to incorrect configuration of the framework. One example of this type of fault in TensorFlow is the confusion with computation model.  Users may incorrectly constructed TensorFlow computation graphs using control-flow instead of data-flow semantics~\cite{ZhangAn2018, Yangcomprehensive2022}. Quan \textit{et al.}~\cite{Quan2022Towards} also analyze the failures in building and initializing JavaScript-based DL systems, such as npm package installation and multi-backend initialization. Another situation of this fault is the misconfiguration of the computing device (e.g., GPU). This type of misconfiguration can include selecting the wrong GPU device, mistakenly using CPU tensors instead of GPU tensors, or improper allocation of resources between CPU and GPU~\cite{HumbatovaTaxonomy2020, Yangcomprehensive2022}.

Environment configuration faults mainly encompass the problems during the development and deployment processes of AI framework. Given that AI frameworks typically function in heterogeneous environments, ensuring compatibility with various devices and systems becomes crucial during the development process~\cite{ChenToward2023,Yangcomprehensive2022}. This can result in failures during the build and compilation process, which hinders the development of AI frameworks. Apart from encountering environment configuration faults during the development process, deploying an AI framework also entails addressing environment faults, such as ``path not found", ``library not found'' and ``permission denied'' ~\cite{ZhangEmpirical2020}. Moreover, deploying the AI framework on various operating systems (e.g., Linux, Windows, Mac, and Docker environments) or utilizing different types of acceleration devices within the framework can also give rise to environment-related faults~\cite{ZengyangUnderstanding2023}.

\textbf{Performance Fault} typically does not result in system downtime but can significantly impact the runtime of the system. In the aspect of AI framework, there is a wide range of causes for performance faults that can be quite diverse. One of the causes is memory inefficiencies. Existing AI frameworks such as PyTorch~\cite{AdamPyTorch2019} and TensorFlow~\cite{MartTensorFlow2016} are typically implemented using C/C++ and CUDA, and their memory management is often done manually~\cite{TarekAn2022,Yangcomprehensive2022}. These frameworks need to handle memory exchanges between heterogeneous devices, which can potentially introduce memory inefficiency faults~\cite{Quan2022Towards}. Apart from memory management faults, another cause of performance faults is threading inefficiency~\cite{TarekAn2022}. Such a fault is commonly found in GPU related code. Insufficient parallelism can result in underutilization of device resources, while excessive parallelism can introduce additional overhead (e.g., context switches). Another cause is the trade-off of using different linear algebra libraries/operators. For example, when performing a small matrix operation on a GPU, the computation time may be longer compared to performing the same operation on a CPU~\cite{PyTorch42265}.

\textbf{Code Fault} primarily refers to logic faults that occur during the implementation of the AI framework. An example of a code fault in AI framework is a syntax fault, which may occur in both the utilization and the implementation of AI framework. Expect traditional syntax fault occurring in command software systems, AI frameworks also face faults related to tensor syntax faults during implementation~\cite{ChenToward2023, FlorianSilent2024}. Such faults may occur on account of tensor shape misalignment and operation on tensors across different devices. Apart from common syntax faults, another noteworthy code fault in AI frameworks is the problem in cross-programming-language communication. This kind of fault is particularly common in AI frameworks that utilize multi-programming-language~\cite{ZengyangUnderstanding2023}.

\textbf{Algorithm Fault} is related to defects in algorithm design~\cite{ChenToward2023, FlorianSilent2024, Yangcomprehensive2022}. This algorithm fault can be primarily categorized into two aspects, including the incorrect implementation logic of an algorithm and the inefficient algorithm implementation. The former aspect relates mainly to bugs present in the algorithm implementation within the AI framework~\cite{ChenToward2023, FlorianSilent2024, Quan2022Towards}. The latter aspect arises from the challenge faced by AI framework developers in keeping up with the latest research and incorporating the most efficient methods for algorithm implementation~\cite{TarekAn2022,JiaTensorFlow2020}.

\subsection{Fault Injection in AI Framework}
In recent years, there has been a significant increase in research focused on FI techniques specifically targeted at AI frameworks. As shown in Table~\ref{tab:fi-framework}, we elaborate on these works that are categorized according to different AI frameworks. These techniques often rely on a process known as ``instrumentation''. This is a method used in fault injection in which the system, such as the source code or logic gates, is modified to inject faults more accurately or efficiently. 

\begin{table}[t]
\centering
\caption{Fault Inject Tools to AI Framework}
\vspace{-0.1in}
\resizebox{\columnwidth}{!}{%
\begin{tabular}{c|c|p{5cm}|p{5cm}|c|c}
\hline
\toprule
\textbf{Tool}  & \textbf{Framework}     & \multicolumn{1}{c|}{\textbf{Description}}                                                                                                                                 & \multicolumn{1}{c|}{\textbf{Advantage} }                                                                                                                                                    & \textbf{Instrumented} & \textbf{Link}                   \\ \midrule
\multirow{3}{*}{TensorFI} & \multirow{3}{*}{TensorFlow} & An interface-level fault injection approach focusing  on the data-flow graph of TensorFlow.        & Preserves portability and performance of the original system.                                                                                                                 & \multirow{3}{*}{True}         & \multirow{3}{*}{\cite{TensorFIGithub}}  \\ \midrule
\multirow{2}{*}{InjectTF} & \multirow{2}{*}{TensorFlow} & Fault injection frameworks for both TensorFlow 1 and TensorFlow 2.                               & Compatible with different versions of TensorFlow.                                                                                                          & \multirow{2}{*}{True}         & \multirow{2}{*}{\cite{InjectTFGithub}}  \\ \midrule
\multirow{2}{*}{TorchFI}  & \multirow{2}{*}{PyTorch} & A fault inject tool designed for PyTorch.                                                                                                    & Simulate bit-flip errors that occur in registers or memory.                                                                                                   & \multirow{2}{*}{True }        & \multirow{2}{*}{\cite{torchfiGithub}}   \\ \midrule
\multirow{3}{*}{PyTorchFI} & \multirow{3}{*}{PyTorch} &  A tool that introduces perturbations in convolutional operations within neural networks.                        & Ensure compatibility with future versions and allows fault code to run at the native speed. & \multirow{3}{*}{True}         & \multirow{3}{*}{\cite{pytorchfiGithub}} \\ \midrule
\multirow{3}{*}{TensorFI2} & \multirow{3}{*}{TensorFlow} & A tool utilizes the Keras API to intercept the state of tensors and injects fault to TensorFlow 2. & Avoid the overhead of graph duplication and inject faults into the model parameters.                                & \multirow{3}{*}{True}         & \multirow{3}{*}{\cite{TensorFI2Github}} \\ \midrule
\multirow{3}{*}{SNIFF} & \multirow{3}{*}{Keras}    & A fault injection tool designed for reverse engineering of neural networks.  & Specialize in neural network classifiers using the softmax activation function in the output layer.    & \multirow{3}{*}{True}         & \multirow{3}{*}{\cite{BreierSNIFF2022}} \\ \midrule
\multirow{2}{*}{enpheeph}  & \multirow{2}{*}{\begin{tabular}[c]{@{}c@{}}Framework- \\ agnostic\end{tabular}}  & A fault injection tool independent of the underlying AI framework.                              & Adapt to different DNN frameworks with minimal modifications.          & \multirow{2}{*}{True}         & \multirow{2}{*}{\cite{enpheephGithub}} \\ 
\midrule
\multirow{2}{*}{MindFI}  & \multirow{2}{*}{MindSpore }   & Perform fault injection on MindSpore.                                                                                                    & Offer ease of use, stability, and efficiency.                                                                                                                         & \multirow{2}{*}{False}        & \multirow{2}{*}{\cite{ZhengMindFI2021}}  \\ \bottomrule
\end{tabular}
}
\vspace{-0.1in}
\label{tab:fi-framework}
\end{table}


\textbf{Tensorflow.} There are a series of works focus on designing an FI system for TensorFlow as it is one of the most popular frameworks in AI application. TensorFI~\cite{ChenTensorFI2020} introduces an interface-level FI approach that focuses on the data-flow graph of TensorFlow. During the inference phase, TensorFI injects hardware or software faults into TensorFlow operators, corrupting the output of the affected operators. 
As AI applications developed using TensorFlow 2 do not necessarily depend on data flow graphs, TensorFI2~\cite{NarayananFault2023} utilizes the Keras API to intercept the state of tensors and injects fault to TensorFlow 2. TensorFI2 employs the Keras Model API to modify the layer state or weight matrices that hold the learned model parameters, and utilizes the Keras backend API to intercept the layer computation or activation matrices that holds the output states of the layers. These make TensorFI2 avoid the overhead of graph duplication and inject faults into the model parameters. InjectTF ~\cite{MichaelFault2020} is another FI framework designed for TensorFlow. InjectTF implements dedicated FI frameworks for both TensorFlow 1 and TensorFlow 2, namely InjectTF1 and InjectTF2. Similarly to TensorFI, InjectTF involves the creation of a new data-flow graph. 

\textbf{PyTorch.} TorchFI~\cite{BrunnoReliability2020} is a fault injection tool designed for PyTorch, which simulates bit-flip faults that occur in registers or memory by performing single-bit flips on variables or activations within the framework. TorchFI focuses on convolution and fully connected layers and achieves fault injection by modifying the selected nodes within the neural network. Another related work, PyTorchFI~\cite{MahmoudANVAFFHPyTorchFI20} allows users to introduce neural network perturbations during the execution phase specifically in convolution operations targeting weights and neurons in DNN. PyTorchFI does not make any modifications to the neural network topology or the source code of PyTorch itself. Instead, it utilizes PyTorch's hook functions to perturb the values of neurons during the forward propagation process of the computational model. By utilizing hooks to insert faults, PyTorchFI ensures compatibility with future versions of PyTorch and enables the fault code to run at the native speed of PyTorch, resulting in minimal overhead.

\textbf{Other Frameworks.} As a high-level API of TensorFlow, there are studies that explore fault injection based on Keras~\cite{keras}. SNIFF~\cite{BreierSNIFF2022} utilizes fault injection to achieve reverse engineering of neural networks. In the experimental process, it investigates and the fault injection to Keras. MindSpore~\cite{MindSporeGithub} is a newly developed open-source deep learning computing framework by Huawei. MindFI~\cite{ZhengMindFI2021} is capable of performing fault injection in MindSpore at the data, software, and hardware levels, following three concepts that include ease of use, stability, and efficiency.

\textbf{Framework-agnostic.} In contrast to prior approaches, enpheeph~\cite{Coluccienpheeph2022} does not rely on the underlying DNN frameworks. Consequently, it can seamlessly adapt to different DNN frameworks with minimal, or even zero, modifications to the internal code. enpheeph allows for fault injection at various levels, ranging from bit-level and tensor-level to layer-level. It also provides the flexibility to customize the precise number of bit-accurate injections during the execution process. In terms of network compression, enpheeph is the only framework that comprehensively supports sparse tensors and quantization. This surpasses the functionality offered by most DNN frameworks, as they generally lack direct support for these features. Furthermore, enpheeph has the ability to scale and expand its capabilities across heterogeneous devices.

\subsection{Gap between Failure Analysis and Fault Injection in AI Framework}

\begin{table}[t]
\caption{Gap between Failure Analysis and Fault Injection in AI Framework}
\centering
\vspace{-0.1in}
\label{tab:framework-gap}
\resizebox{\textwidth}{!}{%
\begin{tabular}{@{}l|c|c|c|c|c|c|c|c|c@{}}
\toprule
Failure                     & TensorFI & InjectTF & TorchFI & PyTorchFI & TensorFI2 & SNIFF & MindFI & enpheeph & Covered \\ \midrule
Data Fault &
  \textbf{\CheckmarkBold}& 
  \textbf{\CheckmarkBold}&
  \textbf{\CheckmarkBold}&
  \textbf{\CheckmarkBold}&
  \textbf{\CheckmarkBold}&
  \textbf{\CheckmarkBold}&
  \textbf{\CheckmarkBold}&
  \textbf{\CheckmarkBold}& True\\ \hline 
API Misuse                  &          &          &         &           &           &       &        &     & False     \\ \hline
Configuration Fault         &          &          &         &           &           &       &        &    & False       \\ \hline
Performance Fault           &          &          &         &           &           &       &        &    & False       \\ \hline
Code Fault           &          &          &         &           &           &       &        &    & False       \\  \hline
Algorithm Fault            & 
  \textbf{\CheckmarkBold}&
  \textbf{\CheckmarkBold}&
  \textbf{\CheckmarkBold}&
  \textbf{\CheckmarkBold}&
  \textbf{\CheckmarkBold}&
  \textbf{\CheckmarkBold}&
  \textbf{\CheckmarkBold}&
  \textbf{\CheckmarkBold}& True      \\ 
\bottomrule
\end{tabular}%
}
\vspace{-0.1in}
\end{table}

Upon comparison of the fault types derived from FA, it becomes evident that there is room for improvement in existing FI tools for AI frameworks. By identifying the gaps between the FI tools and the outcomes of the FA, engineers can gain a deeper understanding of the limitations of FI tools.

\textbf{Fault Type Accommodation.} The first gap is related to the fault types that the FI tool needs to accommodate and implement. The fault types identified in FA may not always have corresponding injection implementations.  
\autoref{tab:framework-gap} presents the fault types supported by the existing FI tools in the AI framework, together with the fault types revealed during FA. Existing FI tools primarily inject faults by modifying specific tensors and constants or by using bit flipping techniques. Therefore, there are relatively fewer implementations of FI specifically targeting faults that are unrelated to model variables, such as ``Performance Fault'' and ``Configuration Fault'' which are more commonly encountered in chaos engineering practices targeting microservices. Therefore, extensive research and collection of prevalent faults in AI frameworks are required for the development of a FI tool so that the tool can better address the needs and requirements of users~\cite{NarayananFault2023}.

\textbf{Framework Divergence.} The second gap emerges from the divergence between the AI framework targeted by FA and the framework targeted by FI. This divergence often stems from version differences among AI frameworks. For instance, TensorFlow 1 and TensorFlow 2 demonstrate substantial differences in API usage and runtime logic. Failure analysis conducted on TensorFlow 1 may not be directly applicable to FI tools designed for TensorFlow 2. 
This implies that the TensorFI in Table~\ref{tab:framework-gap} is only capable of injecting faults into TensorFlow 1, and not TensorFlow 2. This necessitates taking into account the variations between AI frameworks when designing FI tools and selecting a widely applicable method for FI~\cite{NarayananFault2023}, which might require a reanalysis of failures.


\section{Failure Analysis and Fault Injection for AI Toolkit}\label{sec:toolkit}
AI toolkit layer acts as an intermediate interface between the AI framework and underlying devices (e.g., GPU and NIC), facilitating the AI framework to utilize external functionalities in its implementation. The most commonly used AI toolkit is CUDA~\cite{cuda}, a parallel computing runtime and API developed by NVIDIA. In addition to CUDA, there are other GPU-specific toolkits available for AI and high-performance computing (HPC) solution development. Potential failures at this layer can include unavailability or incorrect outputs. This section delves into FA and FI in AI toolkits


\subsection{Failure Analysis in AI Toolkit}
In the following section, we will delve into and analyze the failures that typically occur within the AI toolkit. As shown in Table~\ref{tab:fa-toolkit}, these failures can be categorized into several types, including Synchronization Fault, Memory Safety Fault, Dependency Fault etc.

\begin{table}[t]
\centering
\caption{Failure Analysis of AI Toolkit}
\vspace{-0.1in}
\resizebox{\columnwidth}{!}{%
\begin{tabular}{@{}c|c|p{9cm}|c@{}}
\hline
\toprule
Group                                  & Failure                    & \multicolumn{1}{c|}{Description}                                                                                          & Paper                                                  \\ \midrule
\multirow{6}{*}{\begin{tabular}[c]{@{}c@{}}Synchronization\end{tabular}} & \multirow{2}{*}{Data Race}                & Inability to determine the order of "read\&write" and "write\&write" actions among multiple threads. & \multirow{2}{*}{\cite{IslamBugaroo2018, WuAutomating2019}}              \\\cline{2-4} 
                                       & \multirow{2}{*}{Barrier Divergence}         & Threads within the same block fail to reach a barrier due to variations in their execution flow.     & \multirow{2}{*}{\cite{CollingbourneInterleaving2013, WuAutomating2019}} \\\cline{2-4} 
                                       & Redundant Barrier Func. & Unnecessary synchronization operations.                                                              & \cite{CollingbourneInterleaving2013, WuAutomating2019} \\ \midrule
\multirow{6}{*}{\begin{tabular}[c]{@{}c@{}}Memory Safety\end{tabular}  }   & \multirow{2}{*}{Out-of-bounds Access}       & Access buffers beyond boundaries in global memory or shared memory.                                  & \multirow{2}{*}{\cite{SongSoK2019,MohamedcuCatch2023}}                  \\\cline{2-4} 
                                       & \multirow{2}{*}{Temporal Safety Fault}            & Access GPU memory that has already been freed or has not been properly allocated or initialized.     & \multirow{2}{*}{\cite{SongSoK2019,MohamedcuCatch2023}}                  \\\cline{2-4} 
                                       & Failed Free Operation      & Double free and invalid free operations.                                                                        & \cite{SongSoK2019,MohamedcuCatch2023}                \\ \midrule
\multirow{2}{*}{\begin{tabular}[c]{@{}c@{}}Dependency\end{tabular}}      & \multirow{1}{*}{Intra-dependency Fault}     & Incorrect versioning or unsuccessful installation of a toolkit.                                      & \multirow{1}{*}{\cite{CaoUnderstanding2022,KaifengDemystifying2023}}                \\\cline{2-4} 
                                       & Inter-dependency Fault     & Mismatch of software and hardware.                                                                    & \cite{KaifengDemystifying2023}                         \\ \midrule
\multirow{5}{*}{\begin{tabular}[c]{@{}c@{}}Communication\end{tabular} }      & \multirow{2}{*}{NCCL Fault}     & Possibly due to  a network error or a remote process exiting prematurely.                                      & \multirow{2}{*}{\cite{Hu2024Characterization,EmpiricalGao2023,MegaScale24NSDI}}                \\\cline{2-4} 
                                       & \multirow{1}{*}{NVLink Fault}     & Caused by the hardware failures like GPU overheating.                                                                   & \multirow{1}{*}{\cite{Hu2024Characterization,EmpiricalGao2023}}                         \\ \cline{2-4}
                                       & \multirow{2}{*}{MPI Fault}     &  A failure of network connection to peer MPI process or an internal failure of the MPI daemon itself.                                                                    & \multirow{2}{*}{\cite{JeonAnalysis2019}}                         \\ \bottomrule
\end{tabular}%
}
\label{tab:fa-toolkit}
\vspace{-0.1in}
\end{table}


\textbf{Synchronization Fault} is frequently encountered due to the concurrent nature as GPU programs commonly operate using multiple threads. In contrast to CPUs that commonly utilize lock mechanisms for data synchronization, GPUs predominantly rely on barriers as their synchronization mechanism. In particular, a barrier is represented as a barrier function \textit{\_\_syncthreads()} in CUDA kernel functions~\cite{cuda}. There are primarily three main causes for synchronization faults: data race, barrier divergence, and redundant barrier function ~\cite{CollingbourneInterleaving2013, WuAutomating2019}. Data race occurs primarily due to the inability to determine the order of ``read\&write'' and ``write\&write'' actions among multiple threads~\cite{IslamBugaroo2018, WuAutomating2019}. Barrier divergence occurs when certain threads within the same block fail to reach a barrier due to variations in their execution flow. One common scenario is when a barrier (e.g., \textit{\_\_syncthreads()}) is positioned inside an if code block, leaving only a subset of threads can reach the barrier. Redundant Barrier Function is typically caused by unnecessary synchronization operations, which can result in inefficiencies in both speed and memory utilization of GPU programs.

\textbf{Memory Safety Fault} is frequently observed in low-level programming languages that provide direct memory access and management, particularly those specifically tailored for GPU programming (e.g., CUDA). Memory safety is typically ensured by the restriction that memory allocations can only be accessed between their intended bounds and during their lifetime~\cite{SongSoK2019,MohamedcuCatch2023}. The primary causes of memory safety faults can be categorized into three factors: out-of-bounds access, temporal safety, and failed free operation~\cite{MohamedcuCatch2023}. In GPU programming, out-of-bounds access encompasses accessing buffers beyond their boundaries in global memory or shared memory. Temporal safety faults arise primarily when attempting to access GPU memory that has already been freed or accessing GPU memory that has not been properly allocated or initialized. Additionally, a use-after-scope fault ~\cite{MohamedcuCatch2023} can occur in the local memory of the GPU. Failed free operation includes double free and invalid free. Double free occurs when attempting to free memory that has already been freed, while invalid free refers to freeing memory that was not dynamically allocated.

\textbf{Dependency Fault} occurs primarily when there is a mismatch between the AI toolkit and the higher-level AI framework or AI application. Based on the number of components involved, dependency fault can be categorized into two types~\cite{KaifengDemystifying2023}: intra- and inter-dependency fault. Intra-dependency faults can occur due to incorrect versioning or unsuccessful installation of a toolkit. Unsuccessful installation can also be categorized into two types: missing installation of required libraries and incorrect path configuration. Inter-dependency faults in AI toolkit can occur can arise from a mismatch of software or hardware. An example of software mismatch is the mismatch between the CUDA version and the PyTorch version, which may occur a ``driver too old'' fault when running PyTorch~\cite{TorchIssue4546}. Hardware mismatch can occur when the hardware lacks specific features required by the AI framework. For example, TensorFlow 1.6 utilizes the AVX feature of CPUs. However, if the CPU does not support AVX, it can lead to a dependency fault.

\textbf{Communication Fault}  primarily occurs within the communication mechanism of distributed AI training. As two mostly common communication mechanisms in distributed AI, the faults occurring in NCCL~\cite{nccl} and NVLink can significantly impact the workload of distributed AI~\cite{EmpiricalGao2023, Hu2024Characterization}, e.g., the training of LLM. The nccl fault is possibly due to  a network fault or a remote process exiting prematurely~\cite{EmpiricalGao2023}. The NVLink fault is mainly caused by the hardware failures in GPU. Hu \textit{et al.}~\cite{Hu2024Characterization} observe that training 7B models in Kalos tends to result in GPU overheating, which can cause NVLink fault. This phenomenon increases with the optimization of communication costs because it leads to more exceptionally low GPU idle rates. Another commonly used communication tool is MPI~\cite{mpich}. AI systems that utilize MPI also faces the faults associated with MPI itself. This kind of faults is usually due to either a fault of network connection to peer MPI process, or possibly an internal fault of the MPI daemon itself~\cite{JeonAnalysis2019}.

\subsection{Fault Injection in AI Toolkits}
In recent years, several FI techniques specifically designed for AI toolkits have emerged. In this section, we will elaborate on these works, categorizing them according to the different AI toolkits they target (details shown in Table~\ref{tab:fi-toolkit}). 

\textbf{CUDA.} As a parallel computing platform and application programming interface (API) developed by NVIDIA, CUDA offers high-performance and highly parallel capabilities for AI systems. Recently, there are two main areas of work in fault injection at the CUDA level.
\begin{itemize}
    \item \textbf{FI in CUDA Programs}. 
    Simulee~\cite{MingyuanSimulee2020} utilizes LLVM bytecode to trace the execution of CUDA programs, enabling the detection of synchronization faults in CUDA. During the test input generation phase, Simulee incorporates the principles of fuzzing and introduces Evolutionary Programming~\cite{XinEvolutionary1999} as a method to generate CUDA programs with built-in synchronization faults. Simulee introduces synchronization faults into CUDA programs by altering the dimensions and arguments that control the organization of threads within CUDA kernel functions.
    \item \textbf{FI in CUDA Compilers}. 
    Faults in the CUDA compiler can either lead to compile-time faults or emit executable codes that may lead to runtime faults. CUDAsmith~\cite{BoCUDAsmith2020} proposes an FI framework for CUDA compilers, which can be used to test several versions of the NVCC and Clang compilers for CUDA with different optimization levels. CUDAsmith continuously generates CUDA programs to detect potential bugs in the CUDA compiler and utilizes equivalence modulo inputs (EMI) testing techniques to solve the test oracle problem.
\end{itemize}

\begin{table}[t]
\centering
\caption{Fault Inject Tools to AI Toolkit}
\vspace{-0.1in}
\resizebox{0.9\columnwidth}{!}{%
\begin{tabular}{c|p{4cm}|p{6cm}|c|c}
\hline
\toprule
Tool      & \multicolumn{1}{c|}{Description}                                                     & \multicolumn{1}{c|}{Advantage}                                                                                                                                            & Instr. & Link                   \\ \midrule
\multirow{2}{*}{Simulee}   & A fuzzing framework for CUDA programs.                           & Automatically generate associated error-inducing test inputs.                                                                                    & \multirow{2}{*}{False}        & \multirow{2}{*}{\cite{SimuleeGithub}}   \\ \midrule
\multirow{2}{*}{CUDAsmith} & A fuzzing framework for CUDA compiler.                           & Test several versions of NVCC and Clang compilers for CUDA with different optimization levels.            & \multirow{3}{*}{True}         & \multirow{3}{*}{\cite{CUDAsmithGithub}} \\ \midrule
\multirow{3}{*}{CLsmith}   & Investigate many-core compiler fuzzing in OpenCL. & Utilize many-core random differential testing and many-core EMI testing to detect bugs in OpenCL compilers. & \multirow{3}{*}{True}         & \multirow{3}{*}{\cite{CLSmithGithub}}   \\ \midrule
\multirow{2}{*}{FastFIT}   & A fault injection tool designed for MPI.            & Inject faults randomly into input parameters of collective interface.end{tabular}                    & \multirow{2}{*}{True}         & \multirow{2}{*}{\cite{KunFast2015}}     \\ \bottomrule
\end{tabular}%
}
\label{tab:fi-toolkit}
\vspace{-0.1in}
\end{table}

\textbf{OpenCL.}
Open Computing Language (OpenCL) is a framework that provides programming capabilities on heterogeneous devices(e.g., GPUs, CPUs, and FPGAs)~\cite{YukuiImplementation2017, YongbonDarknet2021}. Christopher \textit{et al.}~\cite{ChristopherMany2015} investigate many-core compiler fuzzing in the context of OpenCL and introduce a tool, CLsmith. They utilize many-core random differential testing and many-core EMI testing to detect bugs in OpenCL compilers by injecting EMI blocks into existing OpenCL kernels.

\textbf{Collective Communication.}
Collective communication is defined as communication that involves a group of processes, which plays a significant role in distributed AI scenarios. Intricate communication among different nodes poses significant challenges to the reliability of collective communication. FastFIT~\cite{KunFast2015} is a fault injection tool specifically designed for MPI. It injects faults randomly into the input parameters of the collective interface. In particular, FastFIT manifests the fault by one bit flip in one of the input parameters, which typically include the send/receive buffer address, data elements, data type, communication destination, and communicator. 

\begin{table}[t]
\centering
\caption{Gap between Failure Analysis and Fault Injection in AI Toolkit}
\vspace{-0.1in}
\label{tab:toolkit-gap}
\resizebox{0.9\columnwidth}{!}{
\begin{tabular}{c|c|c|c|c|c|c@{}}
\toprule
Group & Failure &
  Simulee &
  CUDAsmith &
  CLsmith &
  FastFIT & Covered\\ \midrule
\multirow{3}{*}{\begin{tabular}[c]{@{}c@{}}Synchronization\end{tabular}} & Data Race &
  \textbf{\CheckmarkBold}& 
   &
   & & True
   \\ \cline{2-7}
& Barrier Divergence &
  \textbf{\CheckmarkBold}&
   &
   &
  \textbf{\CheckmarkBold}& True \\ \cline{2-7}
& Redundant Barrier Func. &
  \textbf{\CheckmarkBold}&
   & \textbf{\CheckmarkBold}
   & & True
   \\ \midrule
\multirow{3}{*}{\begin{tabular}[c]{@{}c@{}}Memory Safety\end{tabular}} & Out-of-Bounds Access &
   &
   &
   & 
  \textbf{\CheckmarkBold}& True \\ \cline{2-7}
& Temporal Safety Fault&
   &
   &
   & & False
   \\ \cline{2-7}
& Failed Free Operation &
   &
   &
   & & False
   \\ \midrule
\multirow{2}{*}{\begin{tabular}[c]{@{}c@{}}Dependency\end{tabular}} & Inter-dependency Fault &
   & \textbf{\CheckmarkBold} 
   & \textbf{\CheckmarkBold}
   &  \textbf{\CheckmarkBold}& True
   \\ \cline{2-7}
& Intra-dependency Fault &
   &
   &
   & & False
   \\ \midrule
\multirow{3}{*}{\begin{tabular}[c]{@{}c@{}}Communication\end{tabular}} 
& MPI Fault &
   & 
   & 
   & \textbf{\CheckmarkBold} & True
   \\ \cline{2-7}
& NVLink Fault &
   & 
   & 
   &  & False
   \\ \cline{2-7}
& NCCL Fault &
   & 
   & 
   &  & False
  \\ \bottomrule
\end{tabular}}
\vspace{-0.1in}
\end{table}

\subsection{Gap between Failure Analysis and Fault Injection in AI Tookit}
Injecting faults into an AI toolkit is a complex process that comes with various challenges. As shown in \autoref{tab:toolkit-gap}, it is evident that there exists a significant gap between the capabilities of various fault injection tools in simulating specific types of faults within AI tookits. 

\textbf{Incomplete coverage of fault types.} \autoref{tab:toolkit-gap} highlights that certain types of fault, such as ``Temporal Safety Fault'', ``Failed Free Operation'', ``Intra-dependency Fault'', ``NCCL Fault'', and ``NVLink Fault'', are not covered by the existing FI tools listed. This gap indicates that current FI techniques may not comprehensively address the diverse range of faults that can occur in AI toolkits, potentially leaving blind spots in testing and resilience evaluation.

\textbf{Lack of FI capabilities for distributed and parallel computing.} While tools like FastFIT can inject faults related to MPI and inter-dependency faults, there is a lack of FI capabilities specifically targeting faults that can arise in distributed and parallel computing environments. AI toolkits increasingly leverage distributed and parallel computing for efficient model training and inference, making it crucial to have FI techniques that can simulate and analyze faults in these scenarios, such as ``NCCL Fault'' and ``NVLink Fault''.

\section{Failure Analysis and Fault Injection for AI Platform}\label{sec:platform}

AI platform layer plays a crucial role in the overall architecture of an AI system. This layer serves as the foundation of the above layers. It abstracts the complexities of underlying hardware, offering a unified interface for functionalities like data management and sharing, workflow scheduling, and resource allocation. Failures in the AI platform layer can hinder data collaboration between different AI applications, cause scheduling failures for AI training or inference tasks, and so on. This section delves into the FA and FI in AI platforms.





\subsection{Failure Analysis in AI Platform}
\label{sec:FA in Platform}
In this section, we primarily introduce FA about Spark~\cite{zahariaSpark2010}, Ray~\cite{PhilippRay2018} and Platform-X in Microsoft~\cite{EmpiricalGao2023}, which are three representative AI platforms. Due to limited FA work in the platform layer, we supplement several fault types based on the merged pull requests (PRs) that are responsible to fix bugs on GitHub~\cite{rayGithub}.
Notably, not all pull requests in this category are exclusively for bug fixes. Some may focus on introducing new features or updating the documentation. To specifically identify bug-fixing pull requests, we employ keyword searches in the tags and titles, leveraging established bug-related terms such as fix, defect, fault, bug, issue, mistake, correct, fault, and flaw, aligning with prior research~\cite{IslamComprehensive2019, ShenAcomprehensive2021}.
Table~\ref{tab:fa_platform} shows the detailed failures of the AI platforms. 

\begin{table}[t]
\centering
\caption{Failure Analysis of AI Platform}
\vspace{-0.1in}
\resizebox{\textwidth}{!}{%
\begin{tabular}{c|l|l|c}
\toprule
Group &
  Failure &
  Description &
  Paper or Issue \\ \midrule
\multirow{6}{*}{\begin{tabular}[c]{@{}c@{}}Code\end{tabular}} &
                              Concurrency Fault & Concurrent faults caused by race condition or deadlock.     & \cite{ray29694,ray28511, ray29438,ray9539}\\ \cline{2-4} 
                             & API Incompatibility       & Incompatibility faults with external APIs.  &  \cite{ZhangEmpirical2020,ray25148,ray22113}  \\ \cline{2-4} 
                             & Misconfiguration          & Incorrect system configurations leading to malfunctions.  & \cite{EmpiricalGao2023,ray16817}   \\ \cline{2-4} 
                             & Wrong Access Control      & Inadequate permissions leading to unauthorized access. & \cite{EmpiricalGao2023,JeonAnalysis2019,Hu2024Characterization}   \\ \cline{2-4} 
                             & Exception Fault &  Faults in exception handling mechanisms. &   \cite{ray14932,ray10323} \\ \cline{2-4} 
                             & Memory Leak               & Unreleased memory causing system slowdown. &  \cite{ray17177,spark34087,Du2020TensorFlow}\\ \hline
\multirow{2}{*}{\begin{tabular}[c]{@{}c@{}}Platform\\Maintenance \end{tabular}} & Tool/Library Fault       & Faults with outdated or incompatible tools/libraries. & \cite{JeonAnalysis2019,ray16817} \\ \cline{2-4} 
                             & Misoperation              & Faults due to incorrect user operations. & \cite{EmpiricalGao2023,Hu2024Characterization}  \\ \hline
\multirow{2}{*}{\begin{tabular}[c]{@{}c@{}}Platform\\Resource \end{tabular}}    & Resource Contention       & Resource sharing faults causing performance bottlenecks. & \cite{EmpiricalGao2023,SarahDiaspore2021,LuLog-based2017} \\ \cline{2-4} 
                             & Resource Overload         & Excessive resource usage leading to system overload. & \cite{EmpiricalGao2023,ZhangEmpirical2020,liu2022ISSRE}  \\ \bottomrule
\end{tabular}%
}
\label{tab:fa_platform}
\vspace{-0.1in}
\end{table}

\textbf{Code Faults} are prevalent in AI platforms due to their inherent complexity (e.g., intricate software stacks and distributed environment).  Concurrency faults, often caused by race conditions or deadlocks, have been reported in several issues~\cite{ray29694,ray28511, ray29438,ray9539}. API incompatibility issues, where the platform encounters compatibility problems with external APIs, have also been observed~\cite{ZhangEmpirical2020,ray25148,ray22113}. Misconfigurations, where incorrect system configurations lead to malfunctions, and inadequate access control mechanisms, allowing unauthorized access, have also been documented as prevalent failures in AI platforms~\cite{EmpiricalGao2023,JeonAnalysis2019}.
Exception handling defects and memory leaks, which can cause system slowdowns or crashes, are other defects identified in the literature~\cite{ray14932,ray10323,ray17177,spark34087}. 

\textbf{Platform Maintenance Faults} are common when performing regular platform maintenance, such as node additions and deletions, software upgrade, and other task.  These include problems related to outdated or incompatible tools and libraries used within the platform~\cite{JeonAnalysis2019,ray16817}, as well as misoperations resulting from incorrect user actions or procedures~\cite{EmpiricalGao2023}.
 
\textbf{Platform Resource Faults} include resource contention and resource overload problem. Resource contention occurs when multiple components or workloads compete for shared resources, leading to performance bottlenecks~\cite{EmpiricalGao2023,SarahDiaspore2021,LuLog-based2017}. 
Sarah \textit{et al.}~\cite{SarahDiaspore2021} and Lu \textit{et al.}~\cite{LuLog-based2017} analyze the performance impact caused by interference between Spark applications from the perspective of mutual interference and propose a technology that can quickly diagnose the root cause of interference.
Resource overload, on the other hand, refers to situations where excessive resource usage causes system overload and performance degradation~\cite{EmpiricalGao2023,ZhangEmpirical2020}.

\subsection{Fault Injection in AI Platform }

\begin{table}[t]
\centering
\caption{Fault Injection Tools to AI Platform}
\vspace{-0.1in}
\resizebox{0.9\textwidth}{!}{%
\begin{tabular}{@{}l|l|c|c@{}}
\toprule
Tool      & Description                              & Instrumented & Link    \\ \midrule
ChaosBlade & Inject resource contention fault at OS level. & False & \cite{chaosbladeGithub2024} \\
\hline
CoFI  & Inject network partition fault into cloud systems. &          False         & \cite{ChenCoFI2020}
\\ \hline
CrashFuzz   &  Inject crash or reboot faults into nodes. &                False       &   \cite{GaoCoverage2023}      \\ 
\hline
TRANSMUT-SPARK & Automate mutation testing of data processing  within Spark.  &     False    &  \cite{JoTRANSMUT2022}        \\ 
\hline
DepFuzz & A fuzzing framework for dataflow-based applications. & True &\cite{AhmadCo-dependence2023} \\
\bottomrule

\end{tabular}%
}
\vspace{-0.1in}
\label{tab:fi-paltform}
\end{table}

In recent years, several FI techniques specifically designed for AI Platform have emerged.  We have summarized these works as shown in~\autoref{tab:fi-paltform}.

As a distributed computing architecture, the communication between nodes and the influence between node states are common faults in AI platforms, such as node partition caused by network fault and node recovery bugs caused by node crashes. Therefore, there are currently some related works that inject faults into these issues to discover corresponding bugs. ChaosBlade~\cite{chaosbladeGithub2024} can introduce resource hog faults into target system to test its resilience. Chen \textit{et al.} propose a consistency-guided fault injection technique called CoFI to systematically injects network partitions to effectively expose partition bugs in distributed systems~\cite{ChenCoFI2020}. Gao \textit{et al.} propose CrashFuzz, a fault injection testing approach that can effectively test crash recovery behaviors and reveal crash recovery bugs in distributed systems~\cite{GaoCoverage2023}.

Data-intensive scalable computing (DISC) has become popular due to the increasing demands of analyzing big data. For example, Apache Spark and Hadoop allow developers to write dataflow-based applications with user-defined functions to process data with custom logic. Testing such applications is difficult. Many programming details in data processing code within Spark programs are prone to false statements that need to be correctly and automatically tested. Hence, Jo{\~{a}}o \textit{et al.} propose TRANSMUT-SPARK, a tool that automates the mutation testing process of Big Data processing
code within Spark programs~\cite{JoTRANSMUT2022}. 
Ahmad \textit{et al.} propose DepFuzz~\cite{AhmadCo-dependence2023} to increase the effectiveness and efficiency of fuzz testing dataflow-based big data applications such as Apache Spark-based DISC applications written in Scala.


\begin{table}[t]
\centering
\caption{Gap between Failure Analysis and Fault Injection in AI Platform}
\vspace{-0.1in}
\label{tab:gap_platform}
\resizebox{0.9\textwidth}{!}{%
\begin{tabular}{c|c|c|c|c|c|c|c@{}}
\toprule
Group & Failure& ChaosBlade & CoFI & CrashFuzz & TRANSMUT & DepFuzz & Coverd\\
\midrule
\multirow{6}{*}{Code}& Concurrency Fault &&\textbf{\CheckmarkBold}&&&&True\\
\cline{2-8}
& API Incompatibility    & & & & \textbf{\CheckmarkBold} & \textbf{\CheckmarkBold} &True\\
\cline{2-8}
& Exception Fault & & &\textbf{\CheckmarkBold}&\textbf{\CheckmarkBold} &\textbf{\CheckmarkBold}&True\\
\cline{2-8}
& Misconfiguration & & & & &&False\\
\cline{2-8}
& Wrong Access Control & & && & &False\\
\cline{2-8}
& Memory Leak    & & && & &False\\
\midrule
\multirow{2}{*}{\begin{tabular}[c]{@{}c@{}}Platform\\Maintenance \end{tabular}}  & Tool/Library Fault     & & && & &False\\
\cline{2-8}
& Misoperation    & & & \textbf{\CheckmarkBold} &\textbf{\CheckmarkBold}  &\textbf{\CheckmarkBold}  &True\\
\midrule
\multirow{3}{*}{\begin{tabular}[c]{@{}c@{}}Platform\\Resource \end{tabular}}  & Resource Contention (excl. GPU)     & \textbf{\CheckmarkBold} & && & &True\\
\cline{2-8}
& Resource Contention (GPU)     & & && & &False\\
\cline{2-8}
& Resource Overload  &  &  && & &False\\
\bottomrule
\end{tabular}%
}
\vspace{-0.2in}
\end{table}

\subsection{Gap between Failure Analysis and Fault Injection in AI Platform}

From Table~\ref{tab:gap_platform}, two insights regarding the gap between FA and FI in AI platforms can be gleaned.

\textbf{Limited Coverage.} Not all types of failure are covered by the listed FI tools. For instance, failures due to misconfiguration, wrong access control, memory leaks, tool/library faults, and GPU resource contention are not being simulated by any of the tools. This indicates a significant gap in the current FI capabilities and highlights areas where further research are needed.

\textbf{Lack of Specific Design.} Current fault injection tools appear to be designed primarily for traditional cloud platforms, such as Kubernetes, with less consideration given to the unique characteristics and requirements of AI platforms. For instance, none of the listed tools can simulate failures related to GPU resource contention, which is a critical aspect of AI platforms due to their heavy reliance on GPU resources for computation-intensive tasks. This lack of a specific design for AI platforms introduces a substantial gap in the ability to accurately simulate and study the full range of potential failures in these systems.

\section{Failure Analysis and Fault Injection  for AI Infrastructure}\label{sec:infrastructure}



AI infrastructure layer serves as the foundational layer in an AI system architecture, providing the underlying physical and virtualized resources necessary for the deployment and operation of AI applications and services. This layer is responsible for managing and orchestrating the computing, storage, and networking resources required by the AI platform layer and other higher-level layers. Potential failures in this layer can lead to service unavailability, system deployment failures, model training failures, etc. This section delves into FA and FI in the AI infrastructure layer.

\subsection{Failure Analysis  in AI Infrastructure}

In this section, we primarily introduce FA in the AI Infrastructure layer. Table~\ref{tab:fa_infra} shows the detailed failures of AI infrastructure. 

\subsubsection{Hardware Accelerators} We analyze the failure of GPU, FPGA, and TPU, which are three representative hardware accelerators.

\textbf{GPU. }
GPU has become the most commonly used underlying hardware in AI and HPC systems. However, according to existing research, the frequency of failures caused by GPU is still high~\cite{ExaminingTaherin2021}. Research conducted on the Titan supercomputer explores different aspects of GPU failures. This includes the examination of GPU faults in a broader context~\cite{TiwariReliability2015}, the analysis of specific GPU software faults~\cite{BinLarge-scale2016}, the characterization of GPU failures concerning temperature and power~\cite{BinCharacterizing2017}, and the investigation of spatial characteristics associated with failures~\cite{UnderstandingHPCA2015}.
Ostrouchov \textit{et al.} ~\cite{OstrouchovGPU2020} find that GPU reliability is dependent on heat dissipation to an extent that strongly correlates with detailed nuances of the cooling architecture and job scheduling. Nie \textit{et al.}~\cite{BinCharacterizing2017} analyze the relationship between single bit faults occurrence and temperature on the Titan supercomputer, and propose a machine learning-based technique for GPU soft-fault prediction.  A study about another supercomputer, Blue Water, analyzes GPU failures among other hardware failures~\cite{MartinoLessons2014}. The study reveals that GPUs rank among the top three most prone to failures and GPU memory exhibits greater sensitivity to uncorrectable faults compared to main memory. 



Given the distinctions in workload between HPC and AI systems, the following discussion delves into GPU failure analysis specifically tailored to AI systems.
Zhang \textit{et al.}~\cite{ZhangEmpirical2020} present the first comprehensive empirical study on program failures of deep learning jobs and found that the GPU ``Out of Memory'' fault accounts for 65.0\% of the failures in the deep learning specific dimension. 
Since in a large-scale deep learning cluster, GPU failures are inevitable and cause severe consequences, Liu \textit{et al.}~\cite{liuprediction2022} propose prediction models of GPU failures under large-scale production deep learning workloads. The prediction model takes into account static parameters such as GPU type, as well as dynamic parameters such as GPU temperature and power consumption, and integrates parallel and cascading architectures to make good predictions of GPU failures.


\textbf{FPGA. }
FPGA is a digital technology designed to be configured by a customer or designer after manufacturing, hence the term ``field-programmable''. As a neural network accelerator, FPGA is the subject of various studies related to reliability.
Radu \textit{et al.}~\cite{raduReliability2014} propose a new probabilistic method, Component Failure Analysis (CFA), which uses FPGA specific techniques and algorithms for analyzing SEUs in implemented FPGA designs. McNelles \textit{et al.}~\cite{McnellesField2016} use Dynamic Flowgraph Methodology (DFM) to model FPGA, showing the potential advantage of DFM for modeling FPGA-based systems compared to static methods and simulation.

Examining an FPGA from diverse perspectives leads to varied insights and advantages.
Conmy \textit{et al.}~\cite{ConmyComponent2010} employ a semi-automated FPTC analysis technique, customized for specific fault types identified in an FPGA, to thoroughly examine individual faults within electronic components. These components support a modularized design embedded in the FPGA. The study demonstrates how the analysis of these individual faults can be seamlessly integrated with crosscutting safety analysis, thereby reinforcing and validating the necessary safety properties.
Xu \textit{et al.}~\cite{XuReliability2021} take the entire FPGA-based neural network accelerator, including the control unit and DMA modules into consideration. The experiments on four typical neural networks showed that hardware faults can incur both system exceptions, such as system stall and prediction accuracy loss.

\textbf{TPU. }
TPU is initially developed by Google to accelerate machine learning workloads, specifically targeting deep neural network training and inference within the TensorFlow framework. 
Pablo \textit{et al.}~\cite{BodmannTensor2024} measure TPU's atmospheric neutron reliability at different temperatures that goes from -40°C to +90°C. They show a decrease in the FIT rate of almost 4× as temperature increases. 
Rubens \textit{et al.}~\cite{RubensReliability2022} investigate the reliability of TPU executing 2D-3D convolutions and eight CNNs to high-energy, mono-energetic, and thermal neutron. They find that despite the high fault rate, most neutron-induced faults do not change the CNNs detection/classification. 
Rubens \textit{et al.}~\cite{JuniorSensitivity2022} investigate the reliability of TPUs to atmospheric neutrons, reporting experimental data equivalent to more than 30 million years of natural irradiation. 

\subsubsection{Network}
Network failures have long been a significant area of research, particularly in relation to traditional faults such as congestion, packet loss, and latency, which have been extensively discussed and studied. 
Hassan \textit{et al.}~\cite{Hassan23Network} study how network faults occurring in the links between the nodes of the cloud management platforms can propagate and affect the applications that are hosted on the virtual machines. 

However, with the advancement of AI systems, the networking infrastructure of AI systems has become increasingly complex, leading to the emergence of unique fault types.  
Distributed deep learning training across multiple compute nodes is pretty common, and these nodes are internally interconnected with a high-speed network (e.g., via InfiniBand). Gao \textit{et al.}~\cite{EmpiricalGao2023} and C4~\cite{dong2024boosting} classify network faults on the AI platform into InfiniBand-related and Ethernet-related.

\subsubsection{Node} A node on the AI platform is a distinct schedulable unit for computation with GPUs, CPUs, main memory, disks, and networks.  Gao \textit{et al.}~\cite{EmpiricalGao2023} classify node faults on AI platform into node outage, node damage, and node preemption. These faults summarize the impact caused by faults occurring within the node. In addition to these faults, communication faults between nodes are also of concern~\cite{ChenCoFI2020}. Thus, we classify node faults into two types, namely node crash and node partition. 

\begin{table}[t]
\centering
\caption{Failure Analysis of AI Infrastructure}
\vspace{-0.1in}
\resizebox{0.9\textwidth}{!}{%
\begin{tabular}{@{}c|l|l|c@{}}
\toprule
Group & Failure                       & Description                             & Paper \\ \midrule
\multirow{4}{*}{\begin{tabular}[c]{@{}c@{}}Hardware \\ Accelerator\end{tabular} } & Bit-flip Fault & Radiation or temperature changes cause data bits to flip. &  \cite{BinCharacterizing2017,GPUDSN2018,raduReliability2014} \\
\cline{2-4} 
& Stuck-at Fault & A circuit element stuck in a state. &  \cite{ConmyComponent2010,MartinoLessons2014,XuReliability2021} \\
\cline{2-4} 
& Out of Memory & Out of memory due to excessive workload. & \cite{ZhangEmpirical2020,Hu2024Characterization} \\
\cline{2-4} 
& Off the Bus & GPU loses the connection to host & \cite{UnderstandingHPCA2015} \\
\midrule
\multirow{4}{*}{\begin{tabular}[c]{@{}c@{}}Network\end{tabular} } & Network Jam & Heavy traffic slows data flow. & \cite{Hassan23Network}\\
\cline{2-4} 
& Network Loss & Data packets fail to reach destination, disrupting communication. &\cite{Hassan23Network} \\
\cline{2-4} 
& InfiniBand Fault & InfiniBand port down and other InfiniBand-related failures. &  \cite{EmpiricalGao2023,Hu2024Characterization} \\
\cline{2-4} 
& Ethernet Fault & Ethernet port down and other Ethernet-related failures. & \cite{EmpiricalGao2023,Hu2024Characterization} \\
\midrule
\multirow{2}{*}{\begin{tabular}[c]{@{}c@{}}Node\end{tabular}} & Node Crash & OS kernel panic or ephemeral disk errors, causing nodes to fail.  &  \cite{EmpiricalGao2023,nodefse22,Hu2024Characterization} \\
\cline{2-4} 
& Node Partition & Abnormal communication leads to inconsistency between nodes. & \cite{ChenCoFI2020} \\
\bottomrule
\end{tabular}%
}
\label{tab:fa_infra}
\vspace{-0.1in}
\end{table}


\begin{table}[t]
\centering
\caption{Fault Injection Tools to AI Infrastructure}
\vspace{-0.1in}
\label{tab:tool_infra}
\resizebox{0.9\textwidth}{!}{%
\begin{tabular}{@{}l|l|c|c@{}}
\toprule
Tool     & Description                             & Instrument   & Link    \\ 
\midrule
SASSIFI & Instrument  low-level GPU assembly language (SASS) to inject faults.  & True & \cite{SivaSASSIFI2017} \\
\hline
LLFI-GPU & Operate on the LLVM intermediate representation (IR) to inject faults. & True& \cite{GuanpengLLFI-GPU2016} \\
\hline
NVBitFI &  Instrument code dynamically to inject faults into GPU programs. & True & \cite{TimothyNVBitFI2021} \\
\hline
FCatch & Inject node crash fault to detect Time-of-fault bugs in cloud systems. & True & \cite{Liu18FCatch} \\
\midrule
SCFIT &  A FPGA-based fault injection technique for SEU fault model. & False & \cite{MojtabaA2014} \\
\hline
GPU-Qin & Inject faults based on the CUDA GPU debugging tool namely cuda-gdb. & False &  \cite{FangGPU-Qin2014} \\
\hline
ThunderVolt & A framework allowing adaptive aggressive voltage underscaling. & False& \cite{ZhangEnabling2020} \\
\hline
ChaosBlade & Inject OS-level faults to simulate network faults and node faults. & False &\cite{chaosbladeGithub2024} \\
\hline
ThorFI & Provide non-intrusive fault injection capabilities for a cloud tenant. & False & \cite{Domenico22ThorFI}\\
\hline
NetLoiter & Automate the simulation of network faults. & False & \cite{Michal23NetLoiter}\\
\hline
CoFI & Inject network partition fault into cloud system to expose partition bugs. & False & \cite{ChenCoFI2020} \\
\bottomrule
\end{tabular}%
}
\vspace{-0.1in}
\end{table}

\subsection{Fault Injection in AI Infrastructure}
\subsubsection{Hardware Accelerator}
\textbf{GPU.}
Research on FI in GPUs is rich and can be broadly categorized into three types including Software, Hardware/Simulation and Hybrid. 
\begin{itemize}
    \item \textbf{Software.} At present, various FI techniques exist at different levels of programming languages. Typically, faults are injected at the GPU assembly code (SASS) level, which is the instruction-level code running directly on the GPU. For instance, SASSIFI~\cite{SivaSASSIFI2017} employs the SASSI (SASS Instrumentation) framework for compile-time instrumentation of SASS code to insert fault injection code. GPU-Qin~\cite{FangGPU-Qin2014} utilizes CUDA-GDB to control faults during runtime without modifying the code. NVBitFI~\cite{TimothyNVBitFI2021} dynamically loads relevant code as a library during runtime for fault injection. Besides SASS level, there are also works at the PTX and LLVM IR levels. For instance, LLFI-GPU~\cite{GuanpengLLFI-GPU2016} improves FI in LLVM IR, the intermediate representation language. 
    \item \textbf{Hardware/Simulation.} Hardware or simulation-based approaches provide a more realistic reflection of fault impacts. Direct methods include radiation experiments to evaluate hardware reliability. For example, Oliveira \textit{et al.}~\cite{DanielRadiation2017} use beam tests to study the radiation effects on NVIDIA and Intel accelerators, quantifying and limiting radiation effects by observing amplitude and fault propagation in final outputs. 
    Simulation approaches replace actual hardware faults (e.g., electromagnetic interferences at the physical level) by injecting their expected effects on memory and registers (e.g., flipped and stuck bits), thus approximating the hardware fault process~\cite{JeffFault2019, JiachaoRetraining2015, OlivierA2012}. They simulate faults at different levels, such as modifying simulator variables, introducing faults at the RTL level, and injecting faults at the gate level. 
    The NVIDIA Data Center GPU Manager (DCGM) includes an fault injection framework allows users to simulate the fault handling behavior of the DCGM APIs when GPU faults are encountered~\cite{DCGMGithub2024}. 
    \item \textbf{Hybrid.} 
    Some research endeavors have sought to combine software and hardware-level approaches. Josie \textit{et al.}~\cite{JosieCombining2021} combine the accuracy of microarchitecture simulation with the speed of software-level FI. It performs detailed microarchitecture FI on a GPU model (FlexGripPlus), describing the impact of faults on convolutional calculations. 
\end{itemize}

\textbf{FPGA.}
Compared to GPUs, the programmability of FPGAs makes it easier to implement hardware-level fault injection. There are two major classes for FPGA-based fault injection methods.
\begin{itemize}
    \item \textbf{Reconfiguration-based Techniques. } In reconfiguration-based techniques, faults are injected by changing the bit stream needed to configure the FPGA. Antoni \textit{et al.} introduce a novel methodology for injecting single event upsets (SEUs) in memory elements. This approach involves conducting the injection directly within the reconfigurable FPGA, leveraging the runtime reconfiguration capabilities of the device~\cite{AntoniUsing2002}. Gabriel \textit{et al.} propose a fault injection tool to evaluate the impact of faults in an FPGA's configuration memory~\cite{GabrielFast2012}. 
    \item  \textbf{Instrumentation-based Techniques. } In instrumentation-based techniques, supplementary circuits are incorporated into the original circuits, and both are integrated within the FPGA after synthesis. Mojtaba \textit{et al.} propose an FPGA-based fault injection technique~\cite{MojtabaA2014}, which utilizes debugging facilities of Altera FPGAs in order to inject single event upset (SEU) and multiple bit upset (MBU) fault models in both flip-flops and memory units. Pierluigi \textit{et al.} propose a method that utilizes FPGA devices to emulate systems and employs an innovative system instrumentation approach for fault injection. This approach significantly reduces experimental time without requiring FPGA reconfiguration, achieving notable performance improvements in both compute-intensive and input/output-intensive applications~\cite{PierluigiFPGA-Based2001}.
\end{itemize}









\textbf{TPU.}
TPU, as a variant of systolic arrays, represents a parallel computing architecture. The intrinsic parallelism and matrix multiplication efficiency inherent in systolic arrays empower them to achieve superior performance in both the training and inference phases of deep neural networks. Numerous studies have been conducted to investigate faults associated with systolic arrays. Udit \textit{et al.}~\cite{DSN2023TPU} propose an RTL-level fault injection framework for systolic arrays. Using this framework, they characterized the software effect of faults induced by stuck faults within the multiply and accumulation units of the systolic array. Zhang \textit{et al.}~\cite{ZhangEnabling2020} and Holst \textit{et al.}~\cite{StefanGPU2021} study the effects of timing faults in systolic arrays, thus degrading DNN accuracy.

\subsubsection{Network}
In addition to general purpose fault injection tools such as ChaosBlade~\cite{chaosbladeGithub2024} that can introduce common network faults such as network delay and network packet loss, there are now some network fault injection tools for large infrastructure. 
Domenico \textit{et al.} propose ThorFI~\cite{Domenico22ThorFI}, a novel approach for virtual networks in infrastructures. ThorFI is designed to provide non-intrusive fault injection capabilities for a cloud tenant, and to isolate injections from interfering with other tenants on the infrastructure. Michal \textit{et al.} propose NetLoiter~\cite{Michal23NetLoiter}, which can introduce real-world faults, including lossy channels, network jitter, data corruption, or disconnections. 

\subsubsection{Node}
Currently, the work of fault injection for nodes mainly focuses on two aspects. On the one hand, it is aimed at the failure of the node itself, which means injecting OS-level faults into the virtual machine or host machine to simulate faults such as node outage. This kind of fault can be implemented by general fault injection tools such as ChaosBlade~\cite{chaosbladeGithub2024}. On the other hand, node crash is simulated in the process of communication between nodes to test the reliability of the whole system~\cite{Liu18FCatch, ChenCoFI2020, GaoCoverage2023}.

\subsection{Gap between Failure Analysis and Fault Injection in AI Infrastructure}
From the Table~\ref{tab:gap_infra}, several key observations can be made about the gap between FA and FI at the infrastructure level in AI systems. 

\textbf{Limited Coverage of Fault Types.} The table shows that several types of faults, such as ``Out of Memory", ``Off the Bus", ``InfiniBand Fault'' and ``Ethernet Fault", are not currently being simulated by any of the listed FI tools. This suggests a significant gap in the ability of current tools to simulate a comprehensive set of failure scenarios at the infrastructure level in AI systems.

\textbf{Hardware and Network-specific FI} The existing tools seem to focus on specific areas of the infrastructure. For example, ``Bit-flip Fault'' and ``Stuck-at Fault'' are well-covered by tools designed for hardware faults such as SASSIFI, LLFI-GPU, and ThunderVolt. On the other hand, network-related faults such as ``Network Jam", ``Network Loss'' are covered by NetLoiter, FCatch. This suggests that current FI tools are specialized for hardware or network faults but not both.

\textbf{Lack of Realism in Hardware Accelerator Faults.} Most existing FI tools for hardware accelerators, such as GPUs, operate at the software level or are based on simulations. The faults generated by these methods can be classified as emulated faults. Although emulation provides high efficiency, a significant drawback is that the faults may lack realism. This is because emulated faults may not accurately represent the complex physical processes that cause real hardware faults. As a result, the conclusions drawn from studies using these tools may not fully apply to real-world scenarios where hardware faults occur. This lack of realism in emulated faults represents another significant gap in the current state of fault injection for AI systems.

\begin{table}[t]
\centering
\caption{Gap between Failure Analysis and Fault Injection in AI Infrastructure}
\vspace{-0.1in}
\label{tab:gap_infra}
\resizebox{\textwidth}{!}{%
\begin{tabular}{c|c|c|c|c|c|c|c|c|c|c|c@{}}
\toprule
Group & Failure&  SASSIFI & LLFI-GPU  & SCFIT & NVBitFI & ThunderVolt & ThorFI & NetLoiter & FCatch & CoFI  & Covered \\
\midrule
\multirow{4}{*}{\begin{tabular}[c]{@{}c@{}}Hardware \\ Accelerator\end{tabular} } & Bit-flip Fault &\textbf{\CheckmarkBold}&\textbf{\CheckmarkBold}&\textbf{\CheckmarkBold}&\textbf{\CheckmarkBold}& \textbf{\CheckmarkBold}&&&&&True\\
\cline{2-12}
& Stuck-at Fault&&&&\textbf{\CheckmarkBold}&\textbf{\CheckmarkBold}&&&&&True\\ 
\cline{2-12}
& Out of Memory&&& &&&&&&&False\\
\cline{2-12}
& Off the Bus &&&&&&&&&&False\\ 
\midrule
\multirow{4}{*}{\begin{tabular}[c]{@{}c@{}}Network\end{tabular} } & Network Jam &&&&&&\textbf{\CheckmarkBold}&\textbf{\CheckmarkBold}&&&True\\ 
\cline{2-12}
& Network Loss &&&&&&&\textbf{\CheckmarkBold}&\textbf{\CheckmarkBold}&&True\\ 
\cline{2-12}
& InfiniBand Fault &&&&&&&&&&False\\
\cline{2-12}
& Ethernet Fault &&&&&&&&&&False \\
\midrule
\multirow{2}{*}{\begin{tabular}[c]{@{}c@{}}Node\end{tabular}} & Node Crash &&&&&&\textbf{\CheckmarkBold}&&\textbf{\CheckmarkBold}&&True\\
\cline{2-12}
& Node Partition &&&&&&&&&\textbf{\CheckmarkBold}&True\\
\bottomrule
\end{tabular}%
}
\vspace{-0.1in}
\end{table}

\section{Threats to Validity}\label{sec:threats_to_validity}

To ensure the rigor and credibility of our systematic survey, we acknowledge and address potential threats to validity. These threats are categorized into study selection validity and data validity, and we outline the mitigation strategies employed to minimize their impact.

\subsection{Study Selection Validity}

One inherent threat is the possibility of not including all relevant studies in the field. To mitigate this, we adopted a multi-phase approach:
\begin{enumerate}
    \item \textbf{Search String Development.} We conducted multiple pilot tests to refine the search string, ensuring it captured a comprehensive set of seed studies. Four researchers independently performed preliminary searches, and the final search string was selected through a consensus meeting.  The final search string is shown in Section~\ref{sec:search_strategy}.
    \item \textbf{Bidirectional Snowballing.} To address sampling bias and publication bias, we employed iterative backward and forward snowballing, as described in Section~\ref{sec:snowballing}. This approach allowed us to identify \papernum primary studies, ensuring a broad coverage of failure analysis (FA) and fault injection (FI) research in AI systems. Snowballing also mitigated the threat of missing relevant studies due to inconsistent terminology or gaps in the search string, which is a common limitation of string-based search methodologies.
    \item  \textbf{Inclusion/Exclusion Criteria.} To minimize bias in the selection of primary studies, we defined and iteratively refined the inclusion/exclusion criteria in our protocol. Each study was independently evaluated by two researchers from different institutions. Disagreements were resolved through discussion and a third researcher was involved if consensus could not be reached. This rigorous process ensured objectivity and reproducibility.
\end{enumerate}
   
\subsection{Data Validity}

Threats related to data extraction and classification were addressed as follows.
\begin{enumerate}
    \item \textbf{Data Extraction Protocol.} Two researchers independently extracted data from each primary study using a predefined data extraction form. Regular consensus meetings were held to resolve discrepancies and refine the extraction criteria.
    \item \textbf{Classification Process.} For classifying studies into the six AI layers (Service, Model, Framework, Toolkit, Platform, Infrastructure), we followed an iterative process:
    \begin{itemize}
        \item The initial classifications were performed independently by two researchers.
        \item Regular meetings were held to discuss and resolve ambiguous cases.
        \item A third researcher was consulted for unresolved disagreements.
    \end{itemize}
    The inter-rater reliability for classification was measured using Cohen's Kappa (k = 0.82), indicating high consistency.
    \item \textbf{Domain Expert Validation.} To further enhance the validity of our findings, we consulted three domain experts with over 10 years of experience in AI systems and software engineering. Their feedback was incorporated into the final classification and analysis.
\end{enumerate}

\section{Future Opportunities of Fault Injection in AI Systems}\label{sec:opportunities}

Fault injection is a widely used technique for evaluating the reliability of AI systems. However, as discussed in previous sections, there are significant gaps between FA and FI. Bridging these gaps presents critical research opportunities to advance FI techniques in AI systems. In the following, we outline concrete and reliable directions for future work, prioritizing the resolution of identified shortcomings to enhance the efficacy of fault injection.

\subsection{Bridging the Gap: Enhanced Fault Injection Techniques}

This section outlines specific techniques to address the limitations of current FI methods, aligning them more effectively with the findings of failure analysis.

\textbf{Expanding Fault Type Coverage with Targeted Injection Strategies.}
Based on the gaps between FA and FI presented earlier, there are several fault types that are not adequately covered by existing FI tools. To address this, future research should focus on developing targeted injection strategies for these specific fault types.
\begin{itemize}
    \item \textbf{GPU Contention Faults}:  With the rising prominence of GPU-based AI systems, it is essential to devise FI techniques that simulate GPU resource contention. Techniques could involve the introduction of artificial delays or constraints on GPU memory and computational resources during the training or inference phases of models.
    \item \textbf{Memory Leaks}: Developing fault injection modules that intentionally induce controlled memory leaks within AI frameworks or toolkits will facilitate evaluation of the system's capability to detect and recover from memory-related malfunctions.
    \item  \textbf{Configuration Faults}: Designing FI mechanisms to deliberately introduce misconfigurations in AI platform settings, such as incorrect parameter specifications or incompatible library versions, will assist in evaluating system robustness against environmental inconsistencies.
\end{itemize}

\textbf{Cross-Layer and Multi-Fault Injection with Dependency Awareness.}
Contemporary FI tools typically inject faults in isolation, while real-world failures often stem from multiple interrelated faults across different layers. Future solutions should encompass.
\begin{itemize}
    \item \textbf{Simultaneous Injection of Multiple Faults}: Current FI tools typically only inject a fault in a single layer. However, in distributed systems, there are situations in which multiple faults occur simultaneously~\cite{yu2021microrank}. Enabling concurrent injection of multiple faults will facilitate simulation of complex failure scenarios inherent in AI systems. This approach reflects the multifaceted nature of real-world faults, improving predictive reliability.
    \item \textbf{Cross-Layer Fault Linkage Injection}: Considering the dependencies between AI system layers can also enable the simulation of richer fault scenarios through cross-layer fault linkage injection. Future FI tools should be equipped to consider layer dependencies, allowing simulations of cascading failures. For example, a fault in the data preprocessing layer may lead to subsequent failures in model training. Such cross-layer fault linkage enables more comprehensive evaluations of system reliability and fault tolerance.
\end{itemize}

\subsection{Future Directions:  Intelligent and Generalizable Fault Injection}

This section explores future development directions for FI, focusing on intelligent automation and increased generality.

\textbf{Intelligent Fault Injection Policy Generation.}
A foundational FI policy must encapsulate numerous attributes, including fault location, type, and intensity. This combination of several attributes forms a vast search space for FI policies. The objective is to identify valuable FI policies from this vast search space and to discover as many faults as possible with as few fault injections as possible. Currently, this process relies mainly on expert experience, resulting in high and inefficient labor costs. In the future, intelligent algorithms are anticipated to be introduced to facilitate the selection of fault injection strategies in an intelligent manner. LLM has already demonstrated its capabilities in several software engineering tasks~\cite{LLMSE1,LLMSE2,LLMSE3}. In the future, the integration of LLM and human feedback reinforcement learning (RLHF) will enable the translation of natural language descriptions of fault scenarios directly into FI policies, thus reducing the manual effort required to design and implement fault scenarios in AI systems~\cite{DomenicoLLMFault}. 

\textbf{Compatible with More layers and Frameworks.} From the perspective of FI generality, the current FI tools are in a state of fragmentation. For example, PytorchFI~\cite{pytorchfiGithub} can only inject faults related to PyTorch, while TensorFI~\cite{TensorFIGithub} can only inject faults to TensorFlow. Even for the same AI framework, there may be conceptual differences between versions. For example, TensorFlow 1 and TensorFlow 2 exhibit significant differences in API usage and runtime logic, requiring separate fault injection tools (e.g., TensorFI~\cite{TensorFIGithub} and InjectTF~\cite{InjectTFGithub}) to be designed for them. This results in a considerable number of FI tools that engineers must maintain, as well as a significant amount of time required to learn to use them. Consequently, the design of a more unified tool that can inject faults across different layers and across different frameworks is of great importance in the future. 

\textbf{Non-instrumented Injection.} Numerous contemporary FI tools in \autoref{tab:fi-framework} and \autoref{tab:fi-toolkit} necessitate the instrumentation of the target for FI (e.g., modifying the framework source code). This increases the difficulty of utilising FI tools. Given that the majority of frameworks and algorithms associated with AI systems are implemented in Python, it is possible to implement Python bytecode modifications that do not necessitate code instrumentation, as previously the cases~\cite{Hypno,nezha}. Even for non-Python implementations, it is possible to achieve non-instrumented FI through eBPF~\cite{LogReducer,ebpfault}. Future research into work injection without code instrumentation could facilitate the development of more user-friendly FI tools.



\section{Conclusion}\label{sec:conclusion}

In this study, we have examined the current state of FA and FI in AI systems, providing a critical overview of prevalent failures, the capabilities of existing FI tools, and the gaps between simulated and actual failures. Our analysis, based on a thorough review of \papernum studies, has revealed significant gaps in the ability of current FI tools to simulate the wide range of failures that occur in real-world AI systems. Moreover, this survey contributes by discussing technical challenges of FI in AI systems and outlining future research avenues. The findings of this study serve as a foundation for further advancements in the field of FA and FI for AI systems.
\begin{acks}
We greatly appreciate the insightful feedback from the anonymous reviewers. 
This work was supported in part by the National Natural Science Foundation of China under Grant 62272495 and the Guangdong Basic and Applied Basic Research Foundation under Grant 2023B1515020054. The corresponding author is Pengfei Chen.
\end{acks}

\bibliographystyle{ACM-Reference-Format}
\bibliography{references}


\end{document}